\documentclass{aa}  

\usepackage{graphicx}
\usepackage{txfonts}
\usepackage[colorlinks=true,citecolor=blue,linkcolor=blue,urlcolor=blue]{hyperref}
\usepackage[all]{hypcap}
\usepackage{xcolor}
\usepackage[utf8]{inputenc}
\usepackage{lineno}
\usepackage{color}

\newcommand{\AV}{$A_V$}

\begin{document}

   \title{Diversity and evolution of dust attenuation curves from redshift $\MakeLowercase{z}\sim 1$ to 9}
   \subtitle{}
   \titlerunning{Dust attenuation curves at $\MakeLowercase{z}\sim 1-9$}
   \authorrunning{Shivaei, I. et al.}

   \author{Irene Shivaei \inst{\ref{inst:CAB}}
          \and
          Rohan P. Naidu \inst{\ref{inst:MIT},\ref{inst:NHF}}
          \and
          Francisco Rodr\'{i}guez Montero \inst{\ref{inst:KICP}, \ref{inst:UChicago}}
          \and
          Kosei Matsumoto \inst{\ref{inst:UGent}}
          \and
          Joel Leja \inst{\ref{inst:penn}}
          \and
          Jorryt Matthee \inst{\ref{inst:ista}}
          \and
          Benjamin D. Johnson \inst{\ref{inst:CfA}}
          \and
          Pascal A. Oesch \inst{\ref{inst:UGe}, \ref{inst:DAWN}, \ref{inst:NBI}}
          \and 
          Jacopo Chevallard\inst{\ref{inst:Oxf}}
          \and
          Angela Adamo\inst{\ref{inst:StockU}}
          \and 
          Sarah Bodansky\inst{\ref{inst:UMass}}
          \and 
          Andrew J. Bunker\inst{\ref{inst:Oxf}}
          \and
          Alba Covelo Paz\inst{\ref{inst:UGe}}
          \and 
          Claudia Di Cesare\inst{\ref{inst:ista}}
          \and 
          Eiichi Egami\inst{\ref{inst:UA}}
          \and
          Lukas J. Furtak\inst{\ref{inst:BGU}}
          \and 
          Kasper E. Heintz\inst{\ref{inst:DAWN}, \ref{inst:NBI}}
          \and 
          Ivan Kramarenko\inst{\ref{inst:ista}}
          \and 
          Romain A. Meyer\inst{\ref{inst:UGe}}
          \and 
          Naveen A. Reddy\inst{\ref{inst:UCR}}
          \and
          Pierluigi Rinaldi\inst{\ref{inst:ST}}
          \and 
          Sandro Tacchella \inst{\ref{inst:Cam},\ref{inst:Cavendish}}
          \and 
          Alberto Torralba\inst{\ref{inst:ista}}
          \and 
          Joris Witstok\inst{\ref{inst:DAWN}, \ref{inst:NBI}}
          \and 
          Michael A. Wozniak\inst{\ref{inst:UCR}}
          \and 
          Mengyuan Xiao\inst{\ref{inst:UGe}}
          }

   \institute{
    Centro de Astrobiolog\'{i}a (CAB), CSIC-INTA, Carretera de Ajalvir km 4, Torrej\'{o}n de Ardoz, E-28850, Madrid, Spain \label{inst:CAB}\\
        \email{ishivaei@cab.inta-csic.es}
    \and
    MIT Kavli Institute for Astrophysics and Space Research, Massachusetts Institute of Technology, Cambridge, MA 02139, USA  \label{inst:MIT}
    \and
    NASA Hubble Fellow \label{inst:NHF}
    \and
    Kavli Institute for Cosmological Physics (KICP), University of Chicago, IL 60637, USA \label{inst:KICP}
    \and
    Department of Astronomy \& Astrophysics, University of Chicago, 5640 S Ellis Avenue, Chicago, IL 60637, USA \label{inst:UChicago}
    \and
    Sterrenkundig Observatorium Department of Physics and Astronomy Universiteit Gent, Krijgslaan 281 S9, 9000 Gent, Belgium \label{inst:UGent}
    \and
    Institute for Computational \& Data Sciences, The Pennsylvania State University, University Park, PA 16802, USA \label{inst:penn}
    \and
    Institute of Science and Technology Austria, Am Campus 1, 3400 Klosterneuburg, Austria \label{inst:ista}
    \and
    Center for Astrophysics | Harvard \& Smithsonian, 60 Garden St., Cambridge MA 02138 USA \label{inst:CfA}
    \and
    Department of Astronomy, University of Geneva, Chemin Pegasi 51, 1290 Versoix, Switzerland \label{inst:UGe}
    \and
    Cosmic Dawn Center (DAWN), Denmark \label{inst:DAWN}
    \and
    Niels Bohr Institute, University of Copenhagen, Jagtvej 128, K{\o}benhavn N, DK-2200, Denmark \label{inst:NBI}
    \and
    Department of Physics, University of Oxford, Denys Wilkinson Building, Keble Road, Oxford OX1 3RH, UK \label{inst:Oxf}
    \and
    Department of Astronomy, Oskar Klein Centre, Stockholm University, AlbaNova, SE-106 91 Stockholm, Sweden \label{inst:StockU}
    \and
    Department of Astronomy, University of Massachusetts, Amherst, MA 01003, USA \label{inst:UMass}
    \and
    Steward Observatory, University of Arizona, 933 N. Cherry Ave., Tucson, AZ 85721, USA \label{inst:UA}
    \and
    Department of Physics, Ben-Gurion University of the Negev, P.O. Box 653, Be'er-Sheva 84105, Israel \label{inst:BGU}
    \and
    Department of Physics and Astronomy, University of California, Riverside, 900 University Avenue, Riverside, CA 92521, USA \label{inst:UCR}
    \and
    AURA for the European Space Agency, Space Telescope Science Institute, 3700 San Martin Dr., Baltimore, MD 21218, USA \label{inst:ST}
    \and
    Kavli Institute for Cosmology, University of Cambridge, Madingley Road, Cambridge, CB3 0HA, UK \label{inst:Cam}
    \and
    Cavendish Laboratory, University of Cambridge, 19 JJ Thomson Avenue, Cambridge, CB3 0HE, UK \label{inst:Cavendish}
             \\
                         }

   \date{Accepted for publication in A\&A}
 
  \abstract
   {The UV–optical dust attenuation curve is key to interpreting the intrinsic properties of galaxies and provides insights into the nature of dust grains and their geometry relative to stars. In this work, we constrain the UV-optical slope of the stellar attenuation curve using a spectroscopic-redshift sample of $\sim 3,800$ galaxies at $z\sim 1-9$, to characterize the diversity and redshift evolution of stellar attenuation curves and to gain insight into dust production and evolution at high redshifts. The sample is constructed from three JWST/NIRCam grism surveys in GOODS-S, GOODS-N, and Abell-2744 fields, with a wealth of JWST/NIRCam and HST photometry. With constraints from spectroscopic redshifts and emission line fluxes, we use the Bayesian inference framework of \texttt{Prospector} SED fitting code with a flexible dust model to derive the attenuation curve for each galaxy. We find that the attenuation curve slope varies strongly with {\AV} at all redshifts, becoming flatter at higher attenuation, attributed to the effect of scattering, dust-star geometry, and ISM chemical evolution.
   We find no strong correlation between attenuation curve slope and size or axis ratio, and the trends with stellar mass and star-formation rate are largely driven by their correlation with {\AV}. 
   We find strong evidence that at fixed {\AV}, the curve becomes flatter with increasing redshift. On average, the attenuation curves derived here are shallower than those at $z\sim 0$ and than the SMC curve. The highest redshift galaxies at $z = 7-9$ (124 galaxies -- a significantly larger sample than in previous studies) show slopes even flatter than the Calzetti curve, implying reduced UV obscuration and lower IR luminosities than expected from an SMC dust curve, by as large as an order of magnitude. Hydrodynamical simulations that couple dust growth to gas chemical enrichment successfully reproduce the different loci of high- and low-redshift galaxies in the slope–{\AV} diagram, suggesting that dust in high-redshift galaxies is increasingly dominated by large grains produced in supernova ejecta with limited ISM processing at early times. 
   }


   \maketitle

\section{Introduction}

Dust in the interstellar medium (ISM) of galaxies has long been recognized for its significant attenuation effects, where failing to account for it can lead to significant uncertainties in galaxies' intrinsic properties. To correct the observed light for dust attenuation, one needs to assume a dust attenuation curve, which shows the wavelength dependence of attenuation \citep[see][for a review]{salim20}. The shape of this curve is not universal and depends on both the chemical composition and size distribution of dust grains \citep[e.g.,][]{weingartner01a,jones13,hensley23}, and also the distribution of dust with respect to stars (dust-star geometry; e.g., \citealt{wittgordon00,narayanan18}). The former defines the shape of the \emph{extinction} curve, which quantifies the absorption and scattering along a single line of sight, while both form the \emph{attenuation} curve shape, representing the effective attenuation of integrated emission, incorporating radiative transfer effects due to the spatial distribution of dust and sources.
Galaxy-to-galaxy variations in the extinction or attenuation curves are most pronounced in the UV, which is especially critical for studies of high-redshift galaxies, where observations often capture only the rest-frame UV light. Properly understanding the shape of the attenuation curve is therefore crucial for both its practical application of deriving intrinsic fluxes, and also for setting constraints on the physical properties of dust grains in galaxies.

Decades-long efforts have been dedicated to characterizing the shape of dust curves across redshifts. 
In the Milky Way (MW), Small Magellanic Cloud (SMC), and Large Magellanic Cloud (LMC), extinction curves are derived by comparing observed (dust-obscured) spectra of stars to their well-known intrinsic spectra through various sightlines \cite[e.g.,][]{cardelli89,weingartner01a,gordon03,gordon24}. Beyond the local group, extinction curves can be derived in the afterglow spectrum of Gamma-ray bursts \cite[GRBs; e.g.,][]{corre18,zafar18b,heintz19,heintz23,rakotondrainibe24}, lensed quasars \citep{motta02}, and occulting pairs \citep{keel01}. On the other hand, stellar attenuation curves are more widely studied by comparing the observed spectra or Spectral Energy Distributions (SEDs) from integrated emission of galaxies to assumed intrinsic (dust-free) SEDs \cite[e.g.,][]{calzetti94,calzetti00,wild11,kriek13,reddy15,scoville15,battisti16,battisti17b,battisti17a,salmon16,salim18,buat18,battisti20,shivaei20b,shivaei20a,calzetti21,nagaraj22,markov23,fisher25,markov25a,maheson25}. The main difference between the methodologies used in these studies is in their assumption of the intrinsic SED. Some adopt an SED-fitting approach \citep[e.g.,][]{salmon16,salim18,nagaraj22}, while others compare the SEDs of ``dusty'' galaxies to their low-dust counterparts selected based on similarities in intrinsic properties \citep[e.g.,][]{kriek13,scoville15}, or rely on independent measures of dust attenuation, such as the Balmer decrement, to rank galaxies by dust content \citep[e.g.,][]{calzetti00,reddy15,battisti16,shivaei20a}. Other differences include sample selection and sample size, the use of spectroscopy versus photometry, and the wavelength coverage of the available data. 

The studies of stellar continuum attenuation curves show the significant variation in the shape of attenuation curves (particularly in the UV-optical slope) and many of them explore the main galaxy parameters driving these differences. The goal is twofold: to determine the appropriate attenuation curve for different galaxy types, as this choice can have a strong impact on the physical properties, such as star-formation rates (SFRs), derived from SED fitting \citep{conroy13} or directly from observed fluxes \citep{shivaei20a}, and to gain insights into dust grain properties and the dust-star geometry across various galaxies. Before JWST, many of the studies at $z>0$ were focused on cosmic noon ($z\sim 1-3$).
In particular, some found that the attenuation curves have steeper UV–optical slopes at lower masses \citep{reddy18a} and lower metallicities \citep{shivaei20a,shivaei20b}, which can be indicative of a change in dust grain properties due to different ISM conditions that affect grain growth timescales and/or dust production and destruction processes. At redshifts above cosmic noon, the number of studies was limited before JWST \citep[e.g.,][]{scoville15,cullen17,boquien22,tacchella22}, and hence, an extrapolation of findings at lower redshifts was commonly adopted. This would suggest steeper curves at high redshifts, where galaxies tend to have lower masses and metallicities. This interpretation was further supported by ALMA-based studies of the infrared-excess (IRX) vs. UV spectral slope ($\beta$) diagram, which found steep slopes at high redshifts \citep{bouwens16a,fudamoto20a,fudamoto20b,bowler24}, although some studies reported a wide range of slopes \citep{boquien22}.
With JWST, more progress has been made in characterizing attenuation curves at high redshift directly, including the detection of UV bumps at very early cosmic times indicating fast production of carbonacous dust grains in young galaxies \citep{witstok23,markov23,ormerod25} and estimating the UV-optical slope of the attenuation curve \citep{langeroodi24,markov25a,markov25b,fisher25,mckinney25}. Unlike what was expected from lower-redshift studies, these early studies found, on average, shallow attenuation curves at high redshifts. 

However, differences in sample selection, methodologies, and assumptions about intrinsic SEDs in various studies at various redshifts make it difficult to construct a comprehensive and coherent picture of the diversity and evolution of attenuation curves over cosmic time. What is missing is a study with coherent methodology and sample selection, on a statistically large sample over an extended period in cosmic time. This is the goal of the current work. Having a statistically large sample is crucial for studying average trends, as it reduces the effect of biases from specific sightlines or geometries.
We study the UV–optical slope of the attenuation curve for a sample of $\sim 3800$ galaxies over nearly half the age of the Universe from $z\sim1$ to 9. 
This is the largest complete sample of spectroscopically-confirmed galaxies from three flux-limited JWST/NIRCam grism surveys of All the Little Things \citep[ALT;][]{naidu24}, First Reionization Epoch Spectroscopically Complete Observations \citep[FRESCO;][]{oesch23}, and Complete NIRCam Grism Redshift Survey (CONGRESS; PIs: Egami, Sun).
The sample benefits from a well-understood sample selection function, accurate spectroscopic redshifts from emission lines, and robust nebular emission line measurements from grism spectra and/or a wealth of medium-band filters. Additionally, the sample does not have slit-loss correction uncertainties that can be significant at lower redshifts and is free from the complex selection biases of the slit-spectroscopic sample. The structure of the paper is as follows. Section~\ref{sec:data} describes the sample, observations and data reduction, and the methodology of deriving attenuation curves. Section~\ref{sec:results} covers the main results, including the relation between the slope of the attenuation curve and galaxy parameters, and the average attenuation curves as a function of redshift. In Section~\ref{sec:discussion}, we compare our results with those in the literature from $z\sim 0$ to high redshifts, discuss the observed evolution of the attenuation curve slope in the context of simulations, and the implications for high redshift galaxies. The study and results are summarized in Section~\ref{sec:summary}.
Throughout this paper, we assume a $\Lambda$CDM flat cosmology with $H_0=70$\,km\,s$^{-1}$\,Mpc$^{-1}$ and $\Omega_{\Lambda}=0.7$, and a \cite{chabrier03} initial mass function (IMF). All magnitudes are in the AB system \citep{oke83}.

\section{Data and Method} \label{sec:data}

\begin{figure*}[ht]
        \centering
        \includegraphics[width=.8\textwidth]{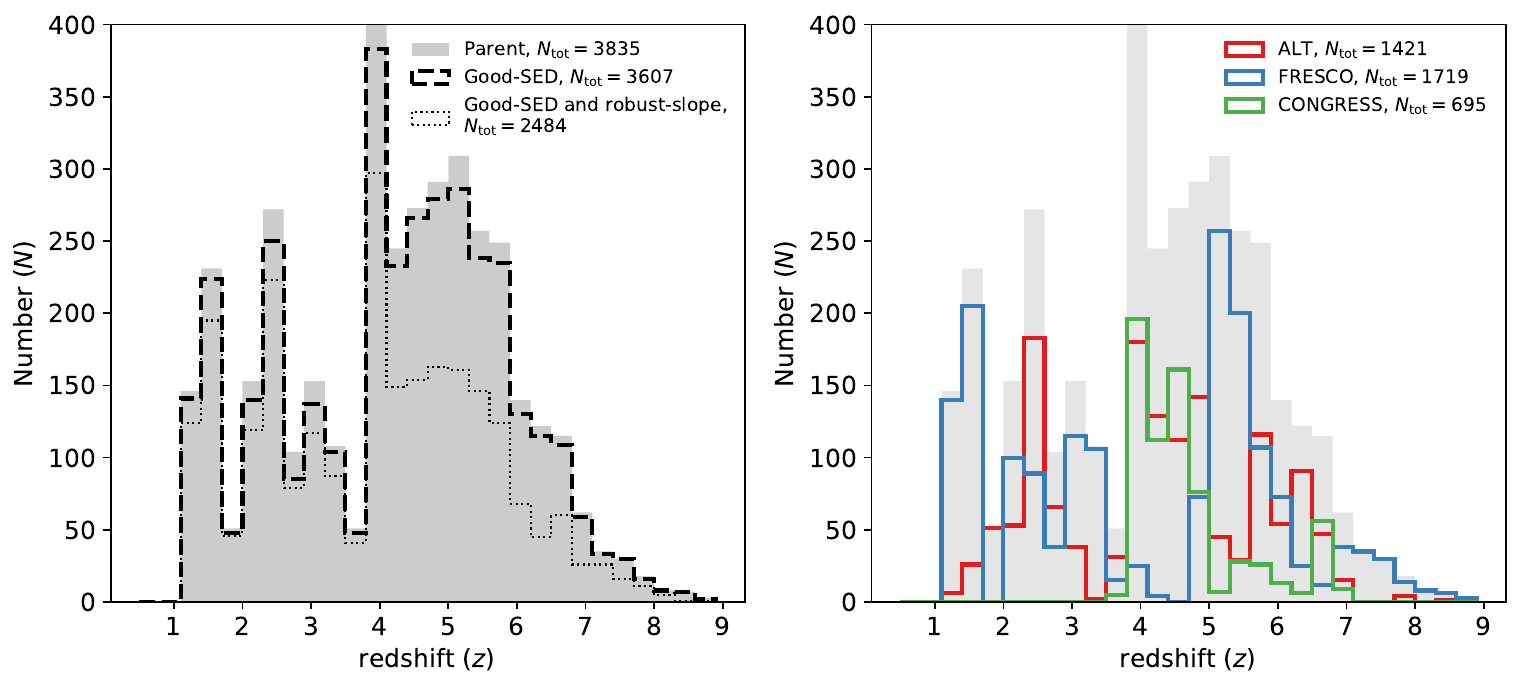} 
    \caption{\label{fig:redshift}
    Redshift distribution of the sample. 
    Left: Redshift distribution of the parent sample is shown with grey filled histogram. Those with good SED fits (``Good-SED'' with reduced $\chi^2 < 2$; 94\% of the parent sample) used here are shown with dashed line. Out of the good-SED sample, those with high-significance attenuation curve slope determination (slope value $<1\sigma$ error, where error is marginalized over all other parameters in the SED modeling) are shown with dotted line (``robust-slope'' sample). The statistical trends throughout the paper are not significantly changed for the parent sample and the good-SED sample, but for accuracy we show the results with the good-SED  sample throughout the paper.
    Right: The sample is further broken into the survey of origin, for all (solid lines) and those with robust slopes and good fits (dashed lines). The redshift peaks correspond to ranges where specific emission lines are shifted into the observed bands (see \S\ref{sec:data}). The sharp peaks in the histogram indicate known overdensities in the fields \citep{helton24,covelo-paz25,herad25}.
    }
\end{figure*}

\subsection{Grism samples}

The sample in this work consists of all spectroscopically-confirmed galaxies at $z>1$ in three large grism surveys of FRESCO \citep[PI: P. Oesch]{oesch23}, ALT \citep[PIs: J. Matthee, R. Naidu]{naidu24}, and CONGRESS (PIs: E. Egami, F. Sun). We refer to the survey papers for details about the survey design and data reduction, briefly described below.
The FRESCO and CONGRESS surveys are observed over 62 arcmin$^2$ of the GOODS-North field,
in F444W (FRESCO) and F356W (CONGRESS) filters. FRESCO also observed 62 arcmin$^2$ of the GOODS-South field in F444W. The grism depth of the two surveys is $2\times 10^{-19}\,\mathrm{erg~s^{-1}~cm^{-2}}$ at $5\sigma$.
ALT is observed in the Abell 2744 (A2744) lensing cluster over a 30 arcmin$^2$ field in the F356W filter, reaching deeper than the other two surveys, with a $5\sigma$ depth of $8\times 10^{-19}\,\mathrm{erg~s^{-1}~cm^{-2}}$ further amplified by lensing.

From the FRESCO survey, we adopted sources with Pa$\alpha$, Pa$\beta$/Pa$\gamma$/Pa$\delta$/He{\sc{i}}, H$\alpha$, and [O{\sc{iii}}]/H$\beta$ emission. The single-line Pa$\alpha$ and H$\alpha$ sources were selected through a line search with a photometric redshift prior, while Pa$\beta$/Pa$\gamma$/Pa$\delta$/He{\sc{i}} and [O{\sc{iii}}]/H$\beta$ sources were selected through a blind search as they have multiple emission lines (see \citealt{meyer24} for more detail). These line emitters correspond to the redshift ranges of $z=1.3-1.66$, $z=2.0-4.2$, $z=4.91-6.59$, $z=6.98-8.95$, respectively and have been presented in \citet{meyer24}, \citet{covelo-paz25}, and \citet{neufeld24}. The Pa$\alpha$ sample extends to $z=1.07$, however, we limit our sample to $z>1.3$ to ensure coverage at $\lambda_{\rm rest}<2000\,\AA$ with at least one photometric band.
The NIRCam/grism data were reduced and processed with the publicly available {\sc Grizli} code\footnote{\url{https://grizli.readthedocs.io/}} \citep{grizli}, and visually inspected. For more details on data reduction and source verification see \cite{covelo-paz25}.
The CONGRESS data was reduced and analyzed in the same way as the FRESCO data \citep{covelo-paz25}. The line search was done for H$\alpha$ and [O{\sc{iii}}]/H$\beta$ over the redshift range of $z=3.78-5.06$ and $z=5.45-6.95$, respectively.
On the other hand, the ALT sample consists of sources from both a blind and a photometric-redshift-prior line search. Therefore, it spans a wider redshift range as it is not limited to the aforementioned emission lines but also includes, e.g., He{\sc{i}}, Pa$\gamma$, Pa$\beta$, H$\gamma$, H$\delta$ emitters. Similar to the FRESCO sample, we limit the ALT sample to $z>1.3$ to ensure sufficient rest-UV coverage with HST and JWST photometric data. For more details on ALT data reduction we refer to \cite{naidu24}.

The redshift distributions of these three surveys are shown in Figure~\ref{fig:redshift}. This is the largest possible {\em complete} sample of spec-$z$ confirmed galaxies at high redshifts in these deep fields, free from complex (and some times random) sample selection biases of slit-spectroscopic surveys. The span of the sample over three deep fields reduces possible cosmic variance uncertainties. The depth of the ALT survey, combined with the lensing magnification, enables selection of very faint sources at high redshifts \citep[as faint as $M_{UV}\approx -15$ at $z\sim 4-7$; see fig. 8 of][]{naidu24}. Furthermore, the wealth of deep NIRCam data in these three fields ensures robust SED modeling. The caveat of this sample is that the selection varies with redshift as it has a fixed line flux limit throughout. It is inevitable that the sample at $z \sim 1-3$ includes more dust-obscured galaxies than the sample at $z\sim 3$, as shown in Figure~\ref{fig:slope_av_params}. The differences between the samples at $z > 3$, however, are not as significant. To mitigate potential selection biases, the main analysis in this work is carried out relative to an appropriate baseline (e.g., in bins of {\AV}), thereby minimizing such biases. 

\subsection{Photometric data}

The JWST+HST imaging reduction and photometric catalogs we use in the GOODS \citep{dickinsongoods03} and A2744 fields are discussed in detail in \citet{weibel24, weibel25}. Briefly, \texttt{Grizli} is used to process the NIRCam data following the steps outlined in \citet{valentino23}. The custom photometric pipeline described in \citet{weibel24} is used to perform photometry (\farcs{16} radius apertures) on images PSF-matched to the F444W filter. Tweaks specific to dealing with the cluster environment in A2744 (e.g., median filtering and aggressive segmentation) are discussed in \citet{naidu24}. Processed HST imaging is incorporated from the Complete Hubble Archive for Galaxy Evolution (CHArGE; \citealt{kokorev22}).

In A2744, the JWST NIRCam imaging arises primarily from the UNCOVER \citep{bezanson24}, MegaScience \citep{suess24}, and ALT \citep{naidu24} surveys along with programs \#2883 (MAGNIF, PI: Sun), \#3538 (PI: Iani), \#2756, PI: Chen \citep[e.g.][]{chen22} and \#3990, PI: Morishita \citep[][]{morishita25}. The archival HST imaging is mainly from the Hubble Frontier Fields program \citep{lotz17, steinhardt20}. Taken together, the galaxies studied in this work typically boast coverage in all 20 JWST NIRCam broad and medium bands complemented by HST ACS and WFC3 imaging providing continuous coverage from $\approx0.4-5\mu$m. The observed fluxes are corrected for magnification using the lensing model presented in \citet{furtak23,price25}. Average magnification of the sample is 3.5 with typical lensing model uncertainty of 1\% \citep{furtak23}. The systematic lensing uncertainty for such low-to-intermediate magnification regime is about 10\% \citep{zitrin15}.

In the GOODS fields, the JWST NIRCam imaging is mainly from the FRESCO \citep{oesch23}, JADES \citep{eisenstein26} and JEMS \citep{williams23} surveys, whereas the HST imaging is primarily from CANDELS \citep{koekemoer11}. While every source we study has coverage in HST bands from 0.4-1.6$\mu$m, the JWST coverage varies across the FRESCO (at least F182M, F210M, F444W) and JADES footprints (all NIRCam broadbands along with F335M, F410M). We have verified that even among the small fraction of sources that are covered only by FRESCO+CANDELS, given the known spectroscopic redshifts and emission line fluxes, the quality of fits is excellent and more than sufficient for the task at hand.

Although MIRI data can provide useful constraints on dust content, particularly at the lower-redshift end of our sample, where long-wavelength MIRI filters probe the mid-IR dust emission \citep{shivaei24a,shivaei24b}, we did not include it in our analysis due to the much shallower depth of the existing MIRI data and also the lack of long-wavelength MIRI photometry for most of the sample.

\subsection{SED fitting} \label{sec:sed-fitting}

Stellar population inference and dust attenuation curve fitting is performed using the \texttt{Prospector} SED fitting code \citep{leja17,leja19,johnson21} closely following the choices outlined in \cite{tacchella22, naidu24}. The SED fitting is done on the photometric data and redshifts are fixed to the grism redshifts. For the ALT sample, the wealth of medium-band photometry in the A2744 field (as explained in \S\ref{sec:data}) provides the necessary constraints for assessing the strength of multiple nebular emission lines (which are not all captured in the grism spectra), and hence, more robust inferred SED parameters such as age, stellar mass, and recent star formation history (SFH). While FRESCO and CONGRESS fields (GOODS fields) have some medium-band photometry, we use emission line fluxes (Pa$\alpha$, H$\alpha$, H$\beta$, [O{\sc iii}]) from slitless grism spectroscopy as constraints in the fitting procedure. This approach of SED fitting with emission line information added provides robust constraints on the stellar population and recent star formation in these galaxies \citep{tacchella23}, while the attenuation curve is mainly constrained by the rest-frame UV data.
Below, the general setup of the SED modeling is described, and in \S\ref{sec:method-dustcurve}, we describe the parametrization of the attenuation curve.

\begin{figure*}[ht]
	\centering
	\includegraphics[width=.8\textwidth]{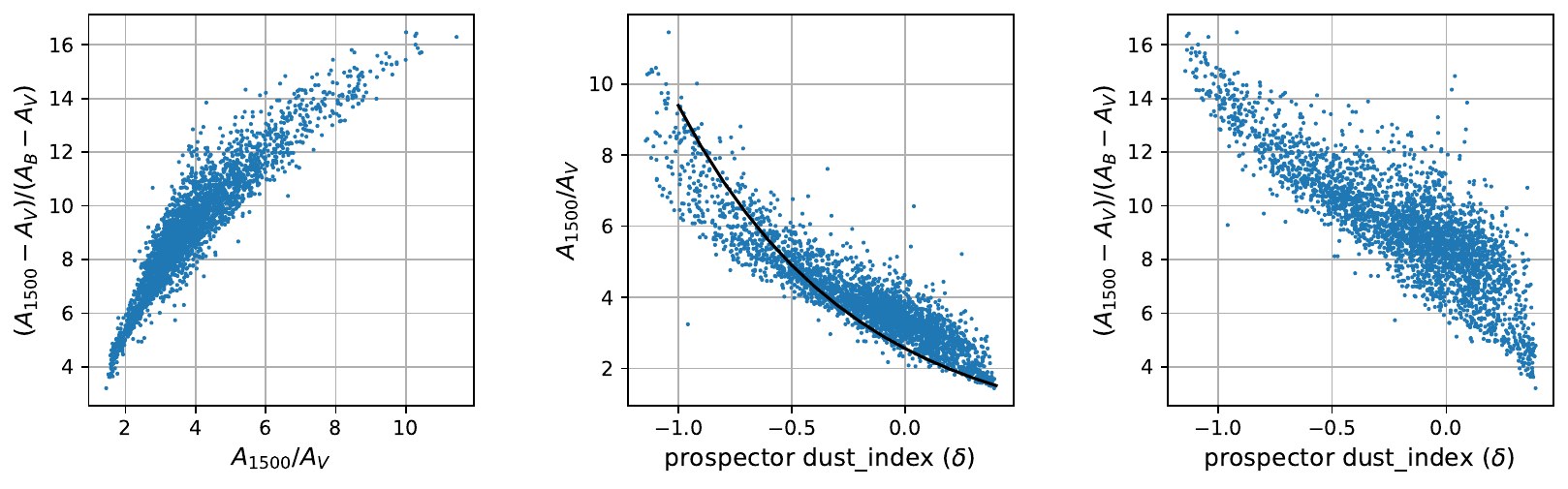} 
	\caption{Comparison of different definitions of attenuation curve slope in the absolute and selective curve formalization, described in \S\ref{sec:slope-def}, and the \texttt{Prospector} power-law index of old stars dust attenuation curve, ``\texttt{dust\_index}'' (see text). In the middle panel, we show the relation between $\delta$ in Equation~\ref{eq:kappa} and the absolute curve slope (discussed later in Equation~\ref{eq:slope-conv}) with a black line. The black curve does not exactly follow the relationship between absolute curve slope and Prospector ``\texttt{dust\_index}'', as the latter is the curve slope only for the old stellar population in the \cite{charlotfall00} parametrization.
	}
	\label{fig:slopes_comp}
\end{figure*}
We use the \texttt{FSPS} \citep[Flexible Stellar Population Synthesis;][]{conroy09} code to model the spectral and photometric properties of galaxies. For building the models, we adopt the \texttt{MIST} \citep[MESA Isochrones and Stellar Tracks;][]{choi17} stellar models, a set of stellar evolution models that provide isochrones and tracks used in population synthesis, and the \texttt{MILES} \citep[Medium-resolution Isaac Newton Telescope Library of Empirical Spectra;][]{MILES2011} spectral library, an empirical stellar spectral library that supplies observed spectra of stars across a range of parameters. We assume a \cite{chabrier03} IMF. Nebular continuum and line emission is modeled with the \texttt{CLOUDY} \citep{ferland17} grid presented in \cite{byler17}. We deploy a non-parametric star formation history and fit for ten bins describing it. The use of a large number of SFH bins is motivated by the extensive dataset covering $0.4-5.0\,\mu$m. Other free parameters are total stellar mass, stellar and gas-phase metallicities, nebular emission parameters, and a flexible dust model (see the section below). For the star formation history, we adopt a ``bursty continuity'' prior \citep{tacchella22}, with the time bins spaced logarithmically up to a formation redshift of $z = 20$. The first two bins are fixed at lookback times of $0-5$ and $5-10$\,Myr to capture bursts responsible for strong emission lines, which are expected to become increasingly common at higher redshifts \citep{boyett24}. 
Uncertainties on SED inferred parameters are obtained from 100 random draws of the \texttt{Prospector} posteriors. 

\subsection{Methodology} {\label{sec:method-dustcurve}}
\subsubsection{Parametrization of dust attenuation curve} 

The attenuation curve is derived assuming 
\begin{equation}\label{eq:dust-attenuation}
	L_{\rm obs} = L_{\rm int} * 10^{-0.4~A_{\lambda}},
\end{equation}
where $A_{\lambda} = E(B-V)~\kappa_{\lambda}$, with $E(B-V)$ being the color excess, $\kappa_{\lambda}$ is in the form of the \cite{calzetti00} curve with a variable slope, $\delta$, following the parametrization of \cite{noll09b} and \cite{salim18}:
\begin{equation}\label{eq:kappa}
	\kappa_{\lambda} = \kappa_{Calz}~\frac{R_V}{R_{V,Calz}}~\left( \frac{\lambda}{5500\AA} \right)^{\delta} + D_{\lambda},
\end{equation}
in which, $\kappa_{Calz}$ is the \cite{calzetti00} attenuation curve (their Equation 4), $R_{V,Calz}$ = 4.05 \citep{calzetti00}, and $D_{\lambda}$ is a Drude profile (a bell-shaped curve that is broader with longer tails compared to a Gaussian profile) for the UV bump \citep{fitzpatrick86}, with a fixed central wavelength (2175\,$\AA$) and width (350\,$\AA$; \citealt{salim18}).
In this parametrization, the flexibility in $\delta$ (\texttt{dust\_index} parameter in our modified \texttt{Prospector} script) accounts for various effects, such as scattering and dust grain properties (size distribution, etc). We note that the variable attenuation curve slope is applied to one of the two dust components of the \cite{charlotfall00} model -  the young stars ($<10\,$Myr) dust attenuation curve has a fixed slope of $-1$, as default. 
However, this assumption means that the output $\delta$ from the SED fitting is not always the same as the slope of the effective attenuation curve at the time of observations, as it corresponds to only one of the dust components. This can be seen in the middle panel of Figure~\ref{fig:slopes_comp}. The effective dust attenuation curve is a combination of the attenuation curves of the two dust components, and depends on the star formation history and age of the galaxy. 

In \texttt{Prospector}, we adopt a uniform prior with minimum (min) and maximum (max) values of $-1$ and 0.4 for \texttt{dust\_index}, and a truncated normal prior with min$=0$, max$=4$, mean$=0.3$, and $\sigma=1$ for \texttt{dust2} (dust optical depth of old stars). Dust optical depth of young stars (\texttt{dust1}) is tied to \texttt{dust2} with a coefficient modeled as a truncated normal distribution with min$=0$, max$=2$, mean$=1$, and $\sigma=0.3$. Finally, the UV bump amplitude is modeled with a uniform prior between 0 and 6, following \cite{salim18}. The UV bump amplitude can be highly uncertain in photometric data fitting, and hence, is not the focus of this work.

To recover the effective stellar dust attenuation curve, we extract the dust-free and nebular emission-free SED, and derive $A_{\lambda}$ by dividing the best-fit observed SED by this intrinsic SED, following Equation~\ref{eq:dust-attenuation}. We emphasize that while the nebular line and continuum emission are modeled in the SED fitting (\S\ref{sec:sed-fitting}), we exclude the nebular component when calculating the {\em stellar} attenuation curve, which is the focus of this work.
For each galaxy, we perform this calculation on the SEDs from 100 random draws of the \texttt{Prospector} posteriors. This ensures that the uncertainties in the derived attenuation-curve parameters (e.g., the slope; \S\ref{sec:slope-def}) reflect the full range of model uncertainties. From these draws, we then compute the median value and the 68\% confidence interval for the parameters of interest.

Two main characteristics of the dust attenuation curve, $\kappa_{\lambda}$, are the shape of the wavelength dependent function, and the normalization value at 5500$\AA$ or $R_V$.
In the absence of IR data to properly anchor the curve through energy-balance arguments, constraining $R_V$ is very uncertain. Therefore, ``non-normalized'' attenuation curve functions are defined that include $R_V$ as a variable. Two of these formalizations that are most common are the selective and absolute attenuation curves, defined as: 
\begin{equation} \label{eq:selective}
    \mathrm{selective~curve:~~}
    \frac{A_{\lambda}-A_{V}}{A_{B}-A_{V}} = \frac{E(\lambda - V)}{E(B-V)} = \kappa_{\lambda} - R_{V}, 
\end{equation}
\begin{equation} \label{eq:absolute}
    \mathrm{absolute~curve:~~}	\frac{A_{\lambda}}{A_V} = \frac{\kappa_{\lambda}}{R_V}.
\end{equation}
In these equations, $A_{\lambda}$ is total attenuation at wavelength $\lambda$, $\kappa_{\lambda}$ is the attenuation curve, $E(B-V)$ is dubbed color-excess, and total attenuation at 5500$\AA$ is $R_{V}$ or $\kappa_{V} = \frac{A_{V}}{E(B-V)}$, the ratio of total to selective extinction in $V$ band. The selective curve formalization provides the ``non-normalized'' selective curve in \citet{calzetti94}. These curves have the same $B-V$ slope and the shape of the curve is independent of the $R_{V}$, making it ideal for studies without IR data.
On the other hand, the absolute curve formalization is more intuitive, but the shape of the curve depends on the normalization ($R_V$). For example, two curves that have the same selective curve slopes but are offset from each other (i.e., different $R_V$) will have different slopes in the absolute formalization of the attenuation curve.
While the selective curve formalization is more appropriate for this study owing to the lack of IR data, we will mostly present the absolute curve results for easier comparison with the literature. 

\subsubsection{Curve slope definition} \label{sec:slope-def}
We define the UV-optical slope, or the ``slope'' in short, of the attenuation curve as the ratio of the curve value at 1500\,$\AA$ to 5500\,$\AA$ (V-band). With this definition, the slope can be calculated for either definitions of the attenuation curve (Equations~\ref{eq:selective},\ref{eq:absolute}). The absolute curve slope is widely adopted in the literature and is commonly called the UV slope of the curve.

The absolute and selective curve slopes closely correlate with each other, as shown in the left panel of Figure~\ref{fig:slopes_comp}. Although the two slopes are derived for the effective attenuation curve, described in the previous Section, they also correlate with the power-law index parameter of the diffuse dust component as well, which is one of the free dust attenuation parameters in our \texttt{Prospector} setup (\texttt{dust\_index} or $\delta$ in Equation~\ref{eq:kappa}; see the previous Section). 
Figure~\ref{fig:slopes_comp} shows a comparison of the two slope definitions with the \texttt{Prospector}'s power-law index parameter of the diffuse dust (\texttt{dust\_index}). Throughout the paper, the effective curves are used to calculate the slope, as the \texttt{Prospector} parameter \texttt{dust\_index} corresponds only to the dust attenuation curve of stars older than 10\,Myr.

\section{Results} \label{sec:results}

\begin{figure*}[ht]
	\centering
	\includegraphics[width=1\textwidth]{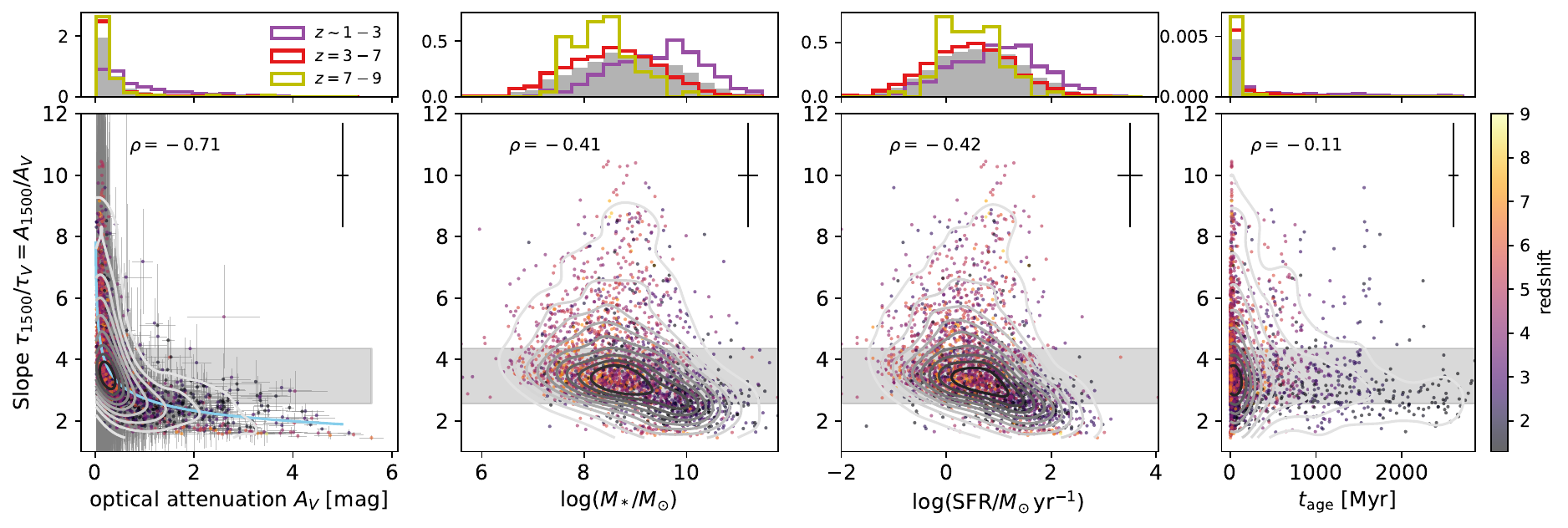} 
	\caption{Absolute attenuation curve slope ($A_{1500\,\AA}$/$A_V$) as a function of galaxy parameters for objects with high-significance slopes (slope value $<1\sigma$ error), from left to right: optical attenuation ({\AV}), stellar mass, SFR averaged over 50\,Myr, and mass-weighted age. Contours represent the average trends of the sample. Individual data points are color-coded by their redshifts. Top histograms show the normalized distribution of the x-axis parameter for the full sample (grey) and in three redshift bins (colored lines).
    We show the uncertainties of individual points in the left panel, and for simplicity only the median uncertainty of the sample in the rest of the panels (crosses in the top-right corner).  These error bars reflect the 16–84th percentiles for the posteriors after marginalizing over all other parameters in the SED modeling.
    The Spearman correlation factors are shown in each panel (all are significant with p-values $<<1$). The strongest relationship is an anti-correlation with {\AV}. A first-order polynomial in log(slope)–log({\AV}) is shown with a blue curve ($\log(\rm{slope}) = -0.23 \times \log(A_V) + 0.44$).
	Grey shaded region marks the range of slopes between those of the Calzetti curve (lower bound) and the SMC curve (upper bound). 
	}
	\label{fig:slope_av_params}
\end{figure*}

The UV-optical slope of the attenuation curve is known to vary significantly from galaxy to galaxy \citep{salim20}, and properly modeling this variation ensures improvement in the accuracy of the inferred galaxy properties across-the-board. However, the drivers behind this variation and whether it correlates with specific galaxy properties are not well understood. In this section, we investigate these questions using our large sample of emission-line selected galaxies across a wide range of redshifts. In \S\ref{sec:slope_gal_parms}, we assess the strength of correlations between the curve slope and galaxy parameters for the full sample. Then, in \S\ref{sec:slope_morph}, we analyze the dependence on morphological parameters for a subset of galaxies. In \S\ref{sec:slope_redshift}, the redshift evolution is explored. In \S\ref{sec:discussion}, we discuss our findings in the context of simulations and other observational studies.

\subsection{Slope vs. galaxy parameters: {\AV}, mass, SFR, age} \label{sec:slope_gal_parms}

Figure~\ref{fig:slope_av_params} shows the curve slope as a function of optical attenuation, {\AV}, stellar mass, SFR, and mass-weighted age, inferred from SED fitting. {\AV} is the total optical attenuation from both dust components assumed in the SED fitting\footnote{{\AV} is tightly correlated with the diffuse dust optical attenuation (derived from the ``\texttt{dust2}'' parameter in \texttt{Prospector}), but often larger (on average 30\%) as it includes the birth-cloud dust component as well.}. 

The slopes of the majority of galaxies lie between the Calzetti and the SMC curve slopes shown with a shaded gray region. The strongest anti-correlation is with {\AV}, with a Spearman correlation factor of $\rho = -0.71$ and a p-value of $<<1$. 
In Appendix~\ref{sec:app-mocks} we show that this relationship is not an artifact of the SED fitting process caused by correlated uncertainties between the two fitted parameters. Although part of this anti-correlation arises from {\AV} appearing on both axes, there are physical explanations for why the two parameters should be related. As {\AV} increases, we observe less deeply into the dust cloud, detecting only the less obscured UV photons. The same effect happens for optical photons as well, however to a lesser degree owing to the wavelength dependence of the dust curve (i.e., one sees deeper into the cloud in optical than in UV, for an optically-thick case).
This results in a shallower attenuation slope, that is, the UV appears less attenuated relative to the optical. Another physical reason for this relationship is that as {\AV} increases and dust-star geometry becomes more mixed, the amount of absorption and scattering increases. The combination of more efficient UV scattering (short-wavelength scattering cross-section is higher), which increases scattering into the line of sight, and the higher forward-scattering of blue photons compared to the more isotropic-scattering of red photons \citep{chevallard13}, which leads to more of the scattered blue photons remaining close to the original direction, leads to a reduced effective UV attenuation, and hence, a shallower attenuation curve. This effect can be seen in the local Universe and in simulations for an edge-on versus a face-on view of the galaxy \citep[e.g.,][]{wild11,chevallard13,matsumoto26}.
The slope-{\AV} anti-correlation has been studied previously in both low- and high-redshift observations \citep[e.g.,][]{salim18,battisti20,nagaraj22,fisher25,markov25b} and also in simulations \citep{wittgordon00,seon16,chevallard13,sommovigo25}. We will compare our results with the literature in \S\ref{sec:slope_redshift}.

In Figure~\ref{fig:slope_av_params}, both stellar mass and SFR show significant anti-correlations with the attenuation curve slope. However, when accounting for their dependence on {\AV} using partial Spearman correlation coefficients, the correlations weaken substantially, reducing to $\rho\sim 0.1$. This suggests that the apparent trends with stellar mass and SFR are largely driven by their underlying correlation with {\AV}, rather than reflecting a direct physical connection to the attenuation curve slope.

The anti-correlation with stellar population age is weaker overall, but there is a noticeable trend at older ages where galaxies tend to have shallower slopes. This is due in part to their higher optical attenuation. In contrast, younger galaxies show a wide range of attenuation slopes, suggesting a more complex or varied dust geometry and grain distribution in these systems.

\subsection{Slope and {\AV} variation with morphology} \label{sec:slope_morph}

\begin{figure*}[ht]
	\centering
		\includegraphics[width=.8\textwidth]{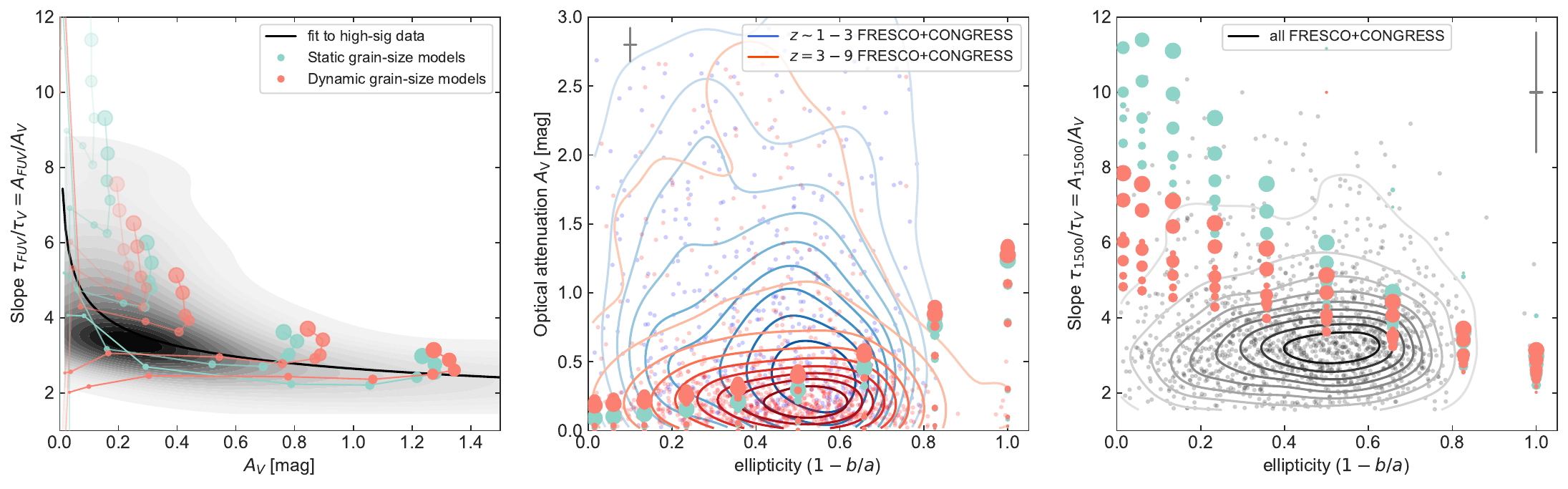} 
    
	\caption{Left: Absolute curve slope versus 5500\,$\AA$ attenuation and for our data with high-significance slope (black contours) and simulations (blue and red tracks). The black curve is a best-fit polynomial to the data (Figure~\ref{fig:slope_av_params}). Red and blue points indicate \cite{matsumoto26} models with and without dust grain size evolution, respectively. Each track shows the time evolution from 50\,Myr to 5\,Gyr (in order of increasing size, the points correspond to 50, 100, 250, 500\,My, 1, 2, 3, 4, 5\,Gyr). The shading of the tracks correspond to inclination angles of 20, 40, 60, 80, and 90$^{\circ}$, such that the darker tracks have lower inclination angles. Middle and right: Optical attenuation {\AV} and attenuation curve slope as a function of ellipticity (derived from axis ratio, $1-(b/a)$) for observations (FRESCO and CONGRESS datasets only; contours and small dots) and simulations (same symbols as the left panel). Assuming thin disks, the ellipticity parameter traces inclination. We divide the observed data to those at cosmic noon (blue) and higher redshifts (red).
    While the simulations show dependence of slope and {\AV} on inclination, there is no significant correlation between the slope or {\AV} and the inclination in the data. Average uncertainties of the measurements are shown with grey crosses in the panels.
	}
	\label{fig:slope_av_models_inclination}
\end{figure*}

\begin{figure*}[ht]
	\centering
	\includegraphics[width=.7\textwidth]{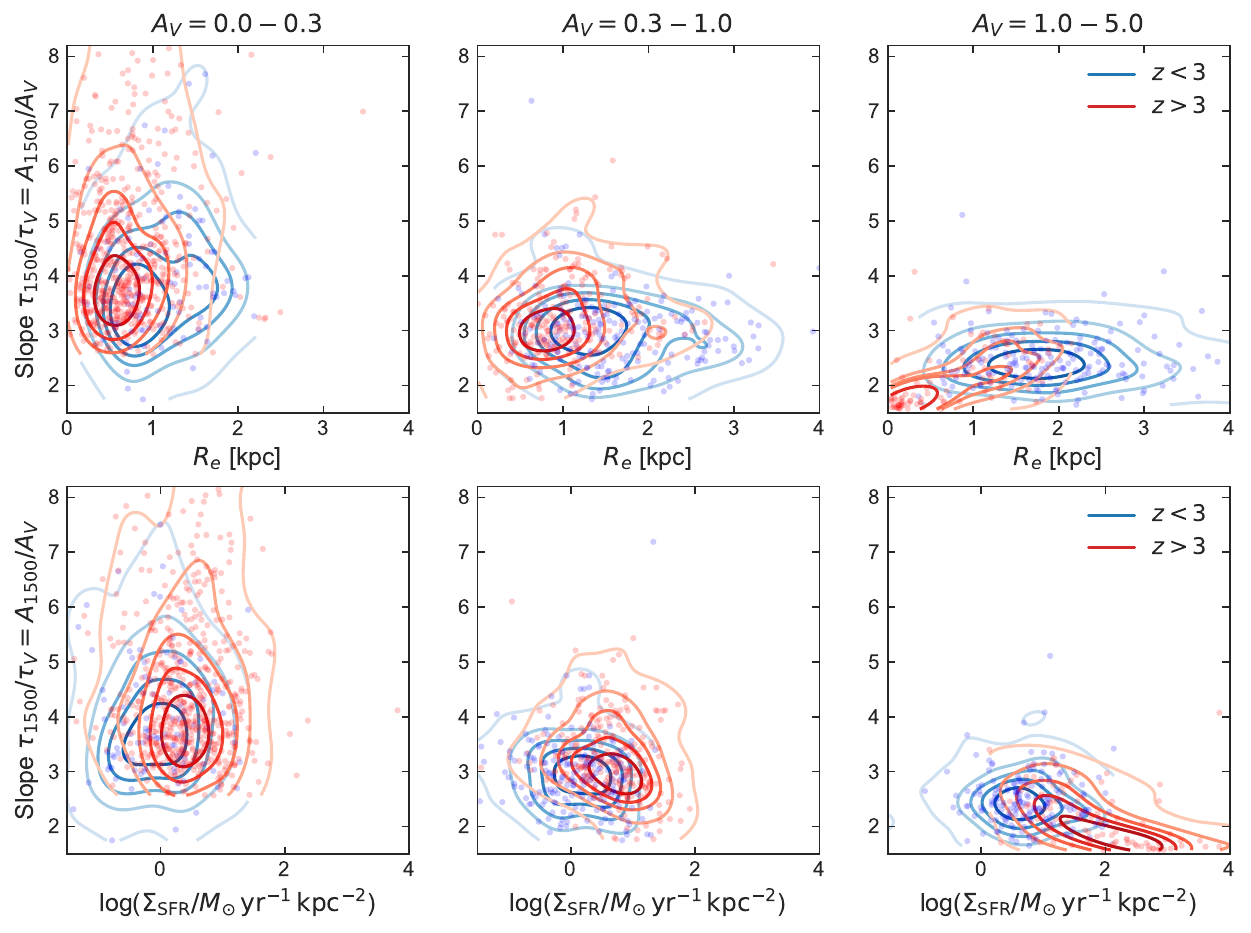} 
    
	\caption{Panels showing variation in the curve slope as a function of effective radius ($R_e$, top row) and SFR surface density ($\Sigma_{\rm SFR}$, bottom row) in bins of {\AV} (columns). Blue and red points and contours show the galaxies below and above $z=3$, respectively (limited to the data in GOODS fields, see text). These panels show that, as expected, not only higher redshift galaxies are smaller with higher $\Sigma_{\rm SFR}$ than lower redshift galaxies, but also, there is a correlation between {\AV} and SFR surface density.
	However, once {\AV} is fixed, the curve slope does not change with respect to either size or SFR surface density -- within each redshift range and across the two redshift bins.
	}
	\label{fig:slope_size_sigma}
\end{figure*}

For nearby galaxies, studies have shown that there is a strong correlation between the attenuation curve slope or dust optical depth with the inclination angle of the galaxies \citep{wild11,battisti17b,maheson24}. This dependence of attenuation parameters on inclination is less clear at higher redshifts, with some studies finding a strong inclination dependence \citep{zuckerman21}, while others do not \citep{lorenz23,maheson24}. Some theoretical studies also show that the slope-{\AV} relation is driven by the inclination effects \citep{faucher24}, while others found the existence of a `quasi-universal' relation between slope and \AV, which is valid for galaxies with intrinsically different dust contents and seen at different inclinations \citep{chevallard13}.
The physical origin of the relation found by \citet{chevallard13} is explained as a combination of geometry and scattering effects. Since red light scatters more isotropically than blue, red photons emitted in the equatorial plane of a galaxy are more likely than blue one to be scattered perpendicularly to the plane and thus escape the galaxy, while blue photons are more likely to be trapped along the plane (scattering is more forward for blue photons) and eventually be absorbed. At low optical depths, this leads to a steepening of the attenuation curve.
At high optical depths, the flattening of the slope originates from the presence of a `mixed' distribution of dust and stars: blue light mostly escapes from regions with optical depth less than unity, i.e, at the edge of the disk, while red photons emerge from deeper layers. This modified the reddening of blue and red light, leading to a flattening of the resulting curve.

\subsubsection{The effect of inclination on dust properties} 

\paragraph{Simulations}
The left panel of Figure~\ref{fig:slope_av_models_inclination} compares the observed slope-{\AV} relation with predictions from the simulation models of \cite{matsumoto26} for an isolated MW-type galaxy.
These models are generated by the \texttt{GADGET4-OSAKA} hydrodynamic code that models the evolution of grain size distributions across $30$ bins ranging from $3.0 \times 10^{-4}$ to $10 \ \mathrm{\mu m}$ and performed post-processing dust radiative transfer calculations with the \texttt{SKIRT} code \citep{baes11,camps15}.
In these models, the evolution of dust mass and grain size distribution is governed by dust production from SNe and AGB stars, dust destruction through gas sputtering and supernova shocks, and interstellar processes such as shattering, coagulation, and accretion \citep{aoyama20,romano22}.
The two models shown in the Figure with different colors have two different dust evolution setups: the static model assumes a constant grain size distribution and composition from \cite{draineli07}, whereas the dynamic model incorporates the time evolution of dust properties. We only use models with scattering effect, as it plays a crucial role in shaping the attenuation curves, particularly at lower inclination angles \citep{matsumoto26} and low optical depths \citep[see fig. 6 of][]{chevallard13}. 
We note that these models do not predict the dust composition in the hydrodynamic simulation -- the dust composition is assigned in post-processing  based on the silicon and carbon abundances \citep{matsumoto24}.
The simulations explore variations in inclination angle, galaxy age, and dust properties that affect the attenuation curves. 

As shown in the left panel of Figure~\ref{fig:slope_av_models_inclination}, both models show a decreasing slope with increasing {\AV}, broadly reproducing the observed locus for the sample galaxies. This behavior can be explained by the effect of scattering, which reduces the effective {\AV} and steepens the attenuation curve at lower optical depths along the line of sight. We discuss the differences between the static and dynamic grain-size models in the context of grain size evolution in high redshift galaxies in \S\ref{sec:discussion}. According to these models, \cite{matsumoto26} show that at low inclination angles as the dust disk becomes more extended (symbols with lighter color/higher transparency in the figure), the attenuation curves steepen after $t_{\rm{age}} = 1$ Gyr due to an increasing impact of scattering. In contrast, at higher inclination angles (darker symbols), the scattering effect is weaker. In these cases, curve steepening is mostly due to the formation of small grains, and/or high fraction of unobscured old stars. 

\paragraph{Observations}
The middle and right panel of Figure~\ref{fig:slope_av_models_inclination} show the comparison of models with ellipticity (defined as $1-(b/a)$, where $b/a$ is the axis ratio) of a subset of our sample. We use the morphological catalog of the DAWN JWST Archive in GOODS fields (i.e., for the FRESCO and CONGRESS samples) from \cite{genin25}. In brief, \cite{genin25} modeled the brightness profile of each source in NIRCam images using the {\sc SourceXtractor++} (SE++), assuming a S{\'e}rsic model \citep{sersic63} to measure the effective radius ($R_e$) and axis ratio ($b/a$). The SE++ tool is performed simultaneously on different bands, yielding morphological values that represent a weighted average over the entire wavelength range used ($0.8-5\,\mu$m). Given the wavelength range, the morphological measures trace the unobscured stellar emission.\footnote{We do not include ALT galaxies in the morphological analysis, as the ALT survey is in a lensing cluster field and the proper investigation of the morphological parameters require careful corrections for magnification, which will be done in a future study.}

We do not find any significant correlation between the curve slope or {\AV} and axis ratio in the observations. The lack of dependence of curve slope on inclination for a sample of galaxies with a wide range of intrinsic dust content has been shown in previous theoretical work \citep{chevallard13}. Particularly, \cite{chevallard13} showed that the curve slope has a strong dependence on the inclination angle only when the sample galaxies have similar intrinsic dust content.

\paragraph{Comparison of simulations and observations}
In the middle and right panels of Figure~\ref{fig:slope_av_models_inclination} we show the trends of the aforementioned \cite{matsumoto26} simulations assuming a thin-disk scenario (i.e., $b/a = \cos(i)$, where $i$ is the inclination angle in the simulations). As these simulations are made for a single galaxy, the effect of inclination on the curve slope is more pronounced than in the observations, which include a large range of intrinsic dust content.
Additionally, the difficulty of modeling the morphological structure of high-redshift galaxies may also play a role in the increased scatter in these relations. The galaxies may not be disks (although the majority of galaxies in this sample have S\'ersic indices $<2$), or the disks may be thick and turbulent, unlike the rotationally-supported disks in the local Universe. This would make morphological modeling challenging and the measured axis ratios would not trace inclination angles. The lack of correlation between dust parameters (stellar and nebular attenuation) with inclination has also been reported before at cosmic noon \citep{lorenz23}. Even in the local Universe, while the inclination has an important effect on the attenuation, it is not the primary parameter that determines the attenuation \citep{maheson24}.

\begin{figure*}[t]
	\centering
	\includegraphics[width=.9\textwidth]{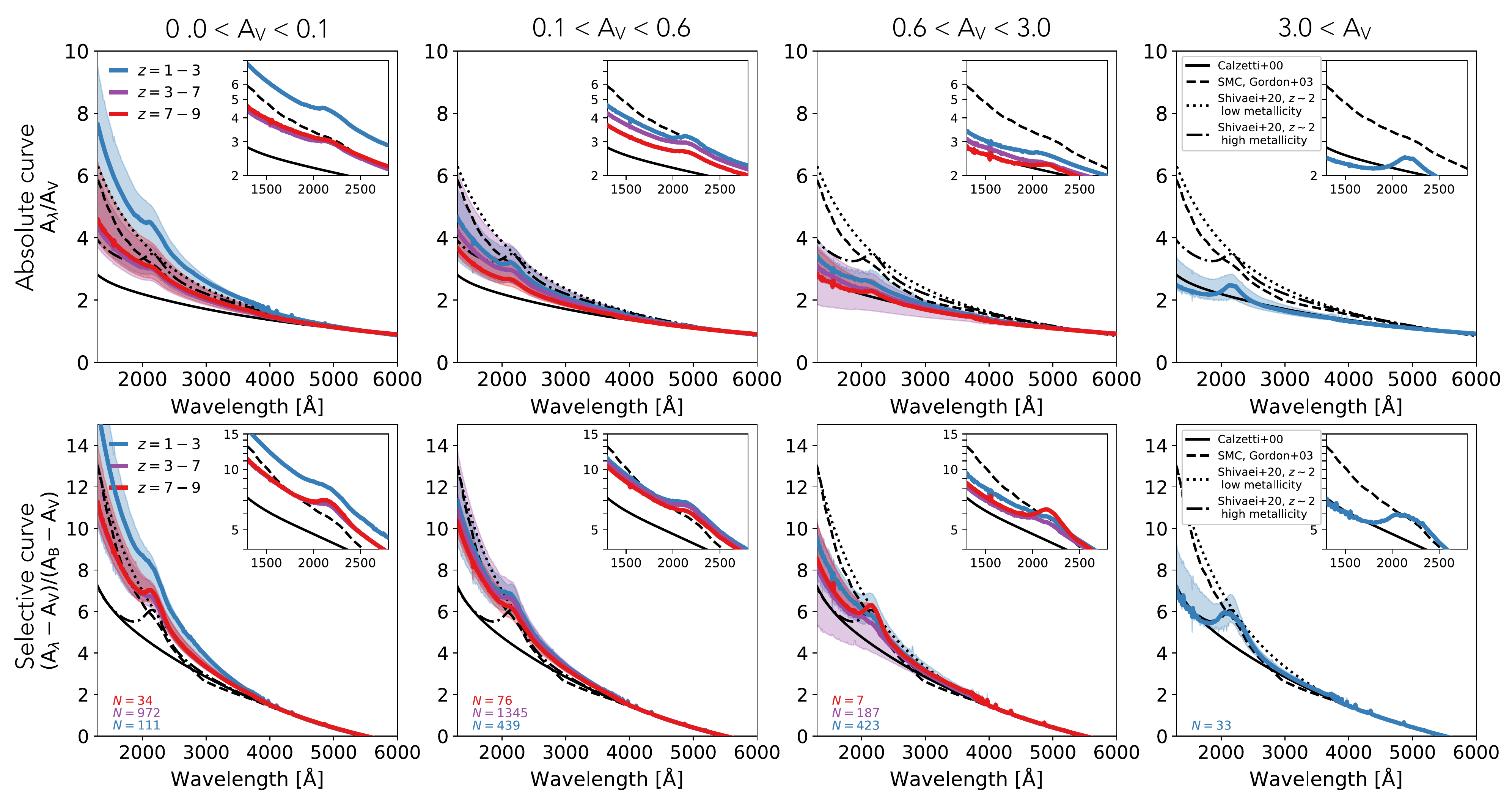}\quad
    
	\caption{Average dust attenuation curves in different {\AV} bins (columns) across multiple redshift ranges (different colors in each panel), for both the absolute (top) and selective (bottom) curve parametrization. The inset panels show a zoomed-in view of the curves in the UV region. Four dust curves from the literature are shown for comparison.
    The average curve in each bin is derived as the weighted 50th percentile of the curves within the bin and the shaded region around each curve represents the 16th to 84th percentiles (see text in Section~\ref{sec:slope_redshift} for more detail). Number of galaxies in each bin is shown in the bottom-left corner. In all redshift ranges, the curve becomes progressively shallower with increasing {\AV}. The $z = 7-9$ curves are consistently shallower than or comparable to those at $z \sim 1-3$. Analytic fits to the absolute curves are given in Table~\ref{tab:curve-fits} and shown in Figure~\ref{fig:attcurve_fits}.}
	\label{fig:attcurve_av}
\end{figure*}

\subsubsection{Size and SFR surface density}
In Figure~\ref{fig:slope_size_sigma}, we explore the effect of size (effective radius, $R_e$) and SFR surface density ($\Sigma_{\rm SFR}$) on the curve slope and {\AV}. Sizes are from \citet{genin25}, as explained in the previous section.
The panels show the variation in slope with respect to size and SFR surface density in bins of {\AV} for FRESCO and CONGRESS sample below and above redshift of $z=3$. Overall, higher redshift galaxies are smaller and have higher $\Sigma_{\rm SFR}$ compared to their lower redshift counterparts, as expected. However, we note that the NIRCam-inferred sizes trace different rest-frame wavelengths at redshifts below and above 3, and we cannot rule out the combined effects of different sizes of young (UV-dominated) vs old (optical-dominated) stellar populations and dust attenuation (affecting the UV light more than optical). Overall, larger $\Sigma_{\rm SFR}$ and smaller size correspond to higher {\AV}, consistent with the expectation that dust optical depth is proportional to dust mass surface density\footnote{Dust optical depth is equal to the production of dust mass absorption coefficient and dust mass density, integrated along the line of sight, which for an evenly distributed dust over a galaxy disk, simplifies to the production of absorption coefficient and dust mass surface density \citep[e.g. see,][]{shapley22}.}. On the other hand, within bins of fixed {\AV}, the attenuation curve slope shows no significant variation with radius or $\Sigma_{\rm SFR}$. These results indicate a lack of observational evidence for dependence of attenuation curve slope on sizes. However, we note that the size measurements are biased preferentially towards unobscured young stellar populations, as at $z<3$, the NIRCam filters trace rest-frame optical to near-IR emission, and at $z>3$ they trace rest-frame UV to optical. A more in-depth modeling of the morphology of galaxies is beyond the scope of this work.

\subsection{Slope and {\AV} variation with redshift} \label{sec:slope_redshift}

In this section, we will focus on the evolution of the slope and attenuation with redshift. Higher redshift galaxies, on average, have lower {\AV} values, as expected. This trend could initially be misinterpreted as evidence for steeper attenuation curves at high redshifts.
Therefore, to assess whether the attenuation curve slope evolves with redshift, potentially due to intrinsic changes in dust grain properties, it is important to control for {\AV}. Figure~\ref{fig:attcurve_av} shows the average attenuation curve of the sample in 4 bins of {\AV}, for 3 redshift ranges. The attenuation curve is shown for both parameterizations: the absolute curve (top row) and the selective curve (bottom row), as described in \S\ref{sec:method-dustcurve}. The average curve in each bin is derived as the weighted 50th percentile of the curves within the bin, where the weights are determined by the uncertainty in the best-fit luminosity at 1600$\AA$ obtained from 100 realizations of each galaxy's fit. The shaded region around each curve represents the 16th to 84th percentiles. Analytic fits to the absolute attenuation curves are provided in Appendix~\ref{sec:curve-fits} and summarized in Table~\ref{tab:curve-fits}.
It is important to keep in mind that the slope of the absolute curve depends on the $R_V$ parameter, which is difficult to constrain in the absence of IR data. Although the shape of the curve differs between the absolute and selective parameterizations, the trends across redshift and {\AV} bins are the same, as described below. 

Figure~\ref{fig:attcurve_av} shows a consistent trend in all redshift bins where the attenuation curve becomes shallower with increasing {\AV}, transitioning from an SMC-like to a Calzetti-like or even shallower curves, in line with the trend shown in Figure~\ref{fig:slope_av_params}. While this trend is evident, the change in slope at $z = 3-7$ between {\AV}$< 0.1$ and {\AV}$= 0.1-0.6$ is relatively modest. Additionally, at fixed {\AV}, the high-$z$ ($z = 7-9$) attenuation curves are always shallower than the low-$z$ ones (at $z \sim 1-3$). 
Notably, the highest-redshift curves lie close to the 16th percentile of the corresponding low-redshift curves across all {\AV} bins, even though the scatter at low-$z$ curve is significant (shaded blue regions). While the intermediate-$z$ ($z \sim 3-7$) curves are often between the high- and low-$z$ ones, the lowest-$A_V$ curve is very similar to the high-$z$ curve. In terms of population properties, the {\AV} distribution within this bin is quite similar across the three redshifts. The only notable difference is that the low-$z$, low-$A_V$ bin has higher average stellar mass and older galaxies compared to the higher-$z$ populations. However, this is not sufficient to explain the very flat attenuation curve at intermediate $z$ at low $A_V$. The scatter in slope values in the lowest {\AV} bin is high, reflecting the large diversity in galaxy dust properties and/or dust–star geometry.
Further exploration of the evolution of high-$z$ attenuation curve slopes at low $A_V$ will be the focus of future work. These results point to possible evolution in the dust properties of galaxies over cosmic time. In the next section, we explore the physical interpretation of these findings and their implications for observations of high-redshift galaxies.

\begin{figure*}[ht]
	\centering
	\includegraphics[width=\textwidth]{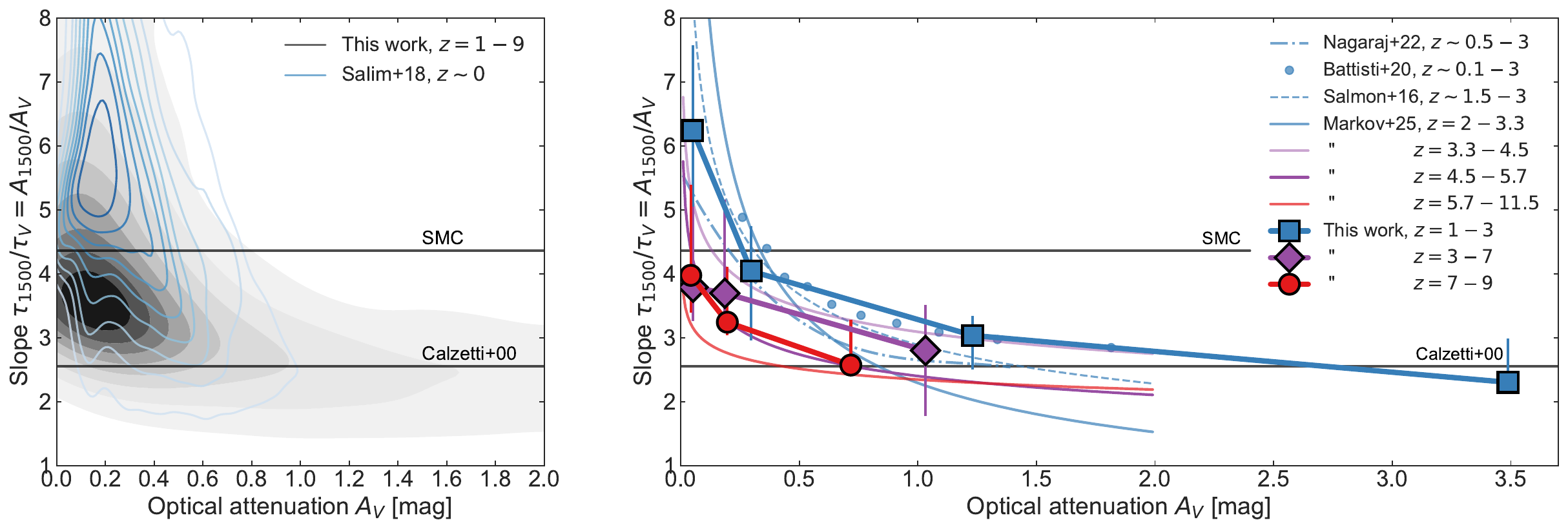} 
	\caption{Slope versus optical dust attenuation in this study compared with literature. Left: Grey contours show restults of this work for the $z\sim 1-9$ sample. Blue contours indicate the star-forming sample of local galaxies from \cite{salim18}, showing steeper attenuation curves at a give {\AV} compared to our sample. The slope of the SMC extinction curve of \cite{gordon03} and the local starburst attenuation curve of \cite{calzetti00} are shown with horizontal lines.
    Right: Results of this work split into three redshift bins and four {\AV} bins shown in Figure~\ref{fig:attcurve_av} (Table~\ref{tab:slope-av-z}). Errorbars reflect galaxy-to-galaxy variations in slope of the attenuation curve in each bin (derived from population percentiles as described in Section~\ref{sec:slope_redshift}). Literature relations are overplotted in matching colors for the corresponding redshift bins: blue for low redshifts, green for medium redshifts, and red for high redshifts. The literature data are from \cite{nagaraj22}, \cite{battisti20}, \cite{salmon16}, and \cite{markov25b}. Our work expands the sample size of high-redshift studies by more than an order of magnitude and reveals a flattening of the attenuation curves at fixed {\AV} toward higher redshifts.}
	\label{fig:slope_av_literature}
\end{figure*}

In Figure~\ref{fig:slope_av_literature}, we put our results in the context of other studies. The figure shows contours for all galaxies with reliable SED fits from this work, along with the values in {\AV} and redshift bins discussed in Figure~\ref{fig:attcurve_av}.
We show the $z\sim0$ contours from \cite{salim18}, who used the CIGALE SED fitting code to derive attenuation curves for a sample of 230,000 galaxies. Additionally, we show the average best-fit curves from several studies covering higher redshifts: $z \sim 0.1-3$ from \citet{battisti20}, $z \sim 0.5-3$ from \citet{nagaraj22}, $z \sim 1.5-3$ from \citet{salmon16}, and $z \sim 2-11$ from \citet{markov25b}. All these works similarly used SED fitting methods to constrain the attenuation curve in bins of {\AV}. 
In all cases, there is an anti-correlation between the slope and {\AV}. A notable trend in the left panel of Figure~\ref{fig:slope_av_literature} is that the $z\sim 0$ contours (blue) lie, on average, above the contours of our study at $z\sim 1-9$ (black). In the right panel, our $z\sim 1-3$ bins (blue squares) are in good agreement with the previous studies at similar redshifts (curves and symbols in blue): the curve slope is relatively steep (even steeper than the SMC curve) at low optical attenuation, and flattens at high attenuation, approaching the Calzetti curve slope. 
At higher redshifts ($z>3$), the average attenuation curve is shallower than the SMC at all {\AV} values. This is in contrast with the results at $z\sim 0$ and the average trend at $z\lesssim 3$. In the highest redshift bin, $z=7-9$, the curve slope is even shallower than the Calzetti curve at high {\AV}. This behavior can be due to a deficiency of small dust grains, a scenario we discuss in comparison with simulations in \S\ref{sec:slope_models}.

\begin{figure}
    \centering
    \includegraphics[width=.9\columnwidth]{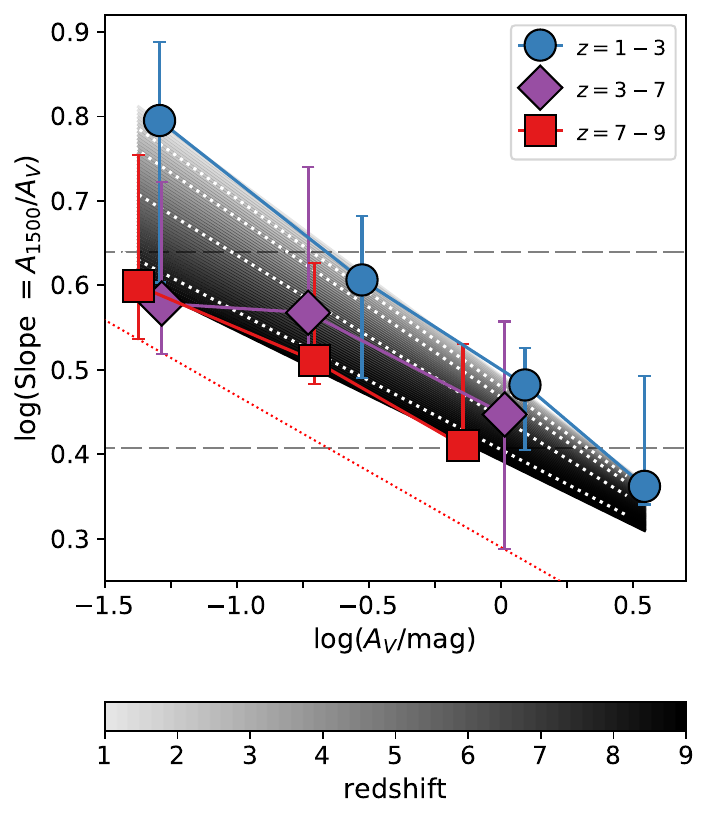}
    \caption{Attenuation curve slope ($A_{1500}$/{\AV}) as a function of {\AV} and redshift (Table~\ref{tab:slope-av-z}). Data binned in {\AV} and redshift are shown with different symbols. The fits, described by Equations~\ref{eq:fit1}-\ref{eq:fit2}, are shown with a black-to-light-gray gradient representing redshifts from $z = 9$ to 1. For clarity $z=2,3,5,8$ curves are shown with white dotted lines, from top to bottom respectively.
    The slope uncertainties reflect the wide variation among the attenuation curve slopes of that bin (derived in the same way as in Figures~\ref{fig:attcurve_av} and \ref{fig:slope_av_literature}). The dotted red line is an estimate of the lower bound of the allowed attenuation curve slope at any redshift (Eq.~\ref{eq:slopemin}). The Calzetti and SMC slopes are shown with horizontal lines.}
    \label{fig:function}
\end{figure}

We provide a functional fit for the attenuation curve slope as a function of {\AV} and redshift. Assuming a linear transition of the slope–{\AV} relationship from the age of the Universe at $z \sim 1$ to that at $z \sim 9$, we adopt the following parameterization:
\begin{equation} \label{eq:fit1}
    \log(\rm{slope})=C_1\times\log(A_V)+C_2,
\end{equation}
where $C_1$ and $C_2$ are linear functions of redshift ($z$):
\begin{equation} \label{eq:fit2}
     C_1=a\times z+a',~~ C_2=b\times z+b'.
\end{equation}
We fit the data in bins of redshift using an ordinary least squares linear regression, yielding the best-fit parameters: $a = 0.010,~a' = -0.242,~b = -0.012,~b' = 0.505$. 
The fit is shown in Figure~\ref{fig:function}. The uncertainty on the slope is 1.08, derived from the average scatter in the attenuation curves of galaxies in each bin shown in Figures~\ref{fig:attcurve_av} and \ref{fig:slope_av_literature}-right. The slope uncertainties are asymmetric and imply a lower bound on the slope, below which the relation is not supported by the data. We show a conservative estimate for this lower bound with a red dotted line in Figure~\ref{fig:function}:
\begin{equation}\label{eq:slopemin}
    \log(\rm{slope_{min}})=-0.179\times\log(A_V)+0.290.
\end{equation}
Here, slope is $(A_{1600}/A_V)$, which can be converted to the $\delta$ in Equation~\ref{eq:kappa}, which is a power-law function of the Calzetti attenuation curve defined in \cite{noll09b} and adopted in other studies \citep{kriek13,salim18} using:
\begin{equation}\label{eq:slope-conv}
    \delta = \frac{\ln(\rm{slope}/2.55)}{\ln(1500/5500)} = \frac{\ln(\rm{slope})-0.94}{-1.30},
\end{equation}
where 2.55 is the slope of the Calzetti attenuation curve.
Equations~\ref{eq:fit1}-\ref{eq:slope-conv} can be implemented in SED fitting codes\footnote{These equations are for the UV-optical attenuation curve. The near-IR part of the curve ($\lambda>6000\AA$) is not studied in this work.}, with truncated priors that enforce the lower bound (Equation~\ref{eq:slopemin}) and allow for steeper curves given the uncertainties discussed above.

\renewcommand{\arraystretch}{1.5} 
\begin{table}[ht]
\caption{Slope and {\AV} values in bins of redshift shown in Figures~\ref{fig:slope_av_literature} and \ref{fig:function}.}
\centering
\begin{tabular}{ccc}
\hline\hline  
redshift range & slope ($\frac{A_{1500}}{A_V}$) & $\langle A_V\rangle$ [mag]  \\
\hline
{$1-3$} & {$6.24^{+1.33}_{-2.75}$} &  {0.05} \\
{} & {$4.04^{+0.70}_{-1.08}$} & {0.30}\\
{} & {$3.03^{+0.31}_{-0.54}$} & {1.23}\\
{} & {$2.30^{+0.69}_{-0.11}$} & {3.49}\\
\hline
{$3-7$} & {$3.79^{+1.25}_{-0.52}$} &  {0.05} \\
{} & {$3.69^{+1.47}_{-0.60}$} & {0.19}\\
{} & {$2.80^{+0.71}_{-1.03}$} & {1.03}\\
{} & {--} & {--}\\
\hline
{$7-9$} & {$3.97^{+1.41}_{-0.58}$} &  {0.04} \\
{} & {$3.24^{+0.87}_{-0.20}$} & {0.20}\\
{} & {$2.57^{+0.71}_{-0.01}$} & {0.71}\\
{} & {--} & {--}\\
\hline
\hline
\end{tabular}
\label{tab:slope-av-z}
\tablefoot{The {\AV} ranges are [0,0.1], [0.1,0.6], [0.6,3], and above 3 for the lowest-$z$ bin only. The average and uncertainties in slopes are derived as explained in Section~\ref{sec:slope_redshift}.}
\end{table}
\renewcommand{\arraystretch}{1.0} 

\begin{figure}[ht]
	\centering
	\includegraphics[width=\columnwidth]{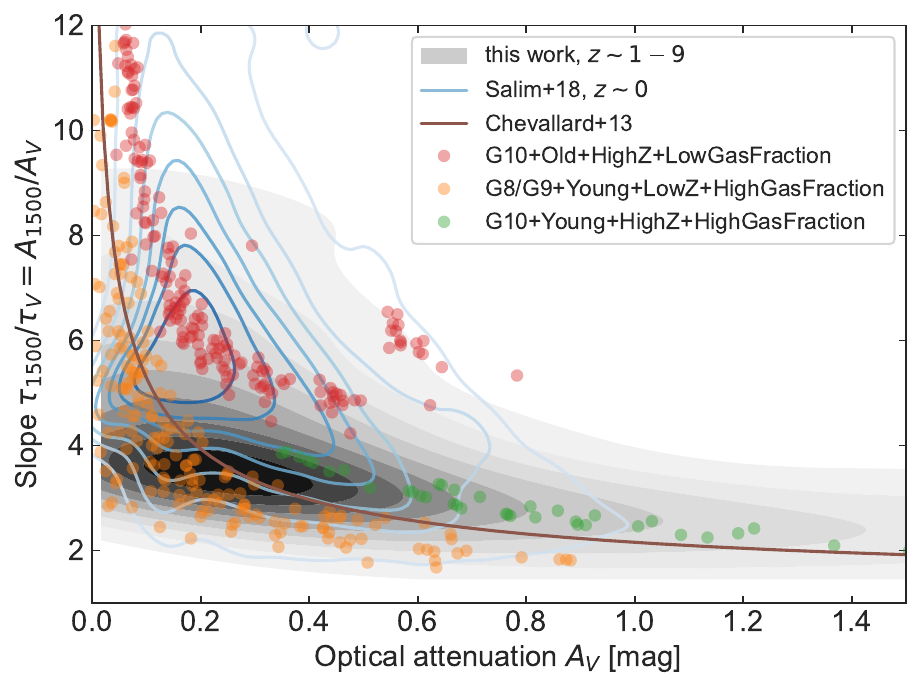} 
	\caption{Comparison of observations with isolated disk galaxy simulations with a range of mass, metallicity, and gas fraction. Grey contours represent the data from this study, and blue contours correspond to local star-forming galaxies from \cite{salim18}. 
    The simulations are adopted from \cite{dubois24} with radiative transfer post-processing. In brief, the G10-Old simulations (red points) model galaxies with large halo masses, high metallicities, older stellar populations and low gas fractions, resembling local universe late-type spirals. The G10-Young simulations (green points) represent massive, starburst galaxies with large halo masses, high metallicities and also high gas fractions. The G8/G9 simulations (orange points) have smaller halo masses, initially low metallicities, and high gas fractions, resembling young galaxies.
    Within each simulation category, individual points show different dust evolution assumptions and initial metallicities. Each point corresponds to a final snapshot after 500\,Myr and has a different (random) inclination angle.
    These simulations reproduce the observed shift of slope–{\AV} trends at different redshifts, supporting that shallow attenuation curves of high-$z$ galaxies at low {\AV} arise from dust mainly produced in stellar ejecta with limited ISM processing. 
    For comparison, we also show the `quasi-universal' relation of \cite{chevallard13} (brown curve), which attributes the slope-{\AV} trend to the effect of scattering and dust-star geometry.
	}
	\label{fig:slope_av_models}
\end{figure}

\section{Discussion} \label{sec:discussion}

\subsection{What drives the slope-{\AV} relation and its redshift evolution?} \label{sec:slope_models}

At any given redshift, both our results and previous studies show a decreasing relation between the UV–optical dust attenuation curve slope and {\AV} (Figure~\ref{fig:slope_av_literature}). This trend can be attributed to the effects of scattering and star–dust geometry \citep[][]{chevallard13,narayanan18,trayford20,matsumoto26}. For example, a non-uniform dust distribution, with a high fraction of optically thin sightlines, can produce flatter slopes. Here, we find that this relation is not universal: high-redshift galaxies tend to shift downward in the slope–{\AV} diagram, exhibiting flatter curves (shallower slopes) at a given {\AV} compared to lower redshift galaxies.

A change in the slope of the attenuation curve can originate from either variations in dust–star geometry \citep[e.g., a non-uniform dust distribution produces flatter curves;][]{wittgordon00} or in the dust grain size distribution \citep[e.g., larger grains yield flatter curves;][]{weingartner01a}. This implies that, {\em at a fixed {\AV}}, high-redshift galaxies should have either a more non-uniform dust–star geometry, larger dust grains, or both, compared to their lower-redshift counterparts.
Studies of nebular dust attenuation suggest non-unity dust covering fractions at cosmic noon \citep{shivaei20a,lorenz25,reddy26}, which might also be applicable to the stellar continuum attenuation. Also, the ALMA observations of cold dust in massive bright galaxies at $z\sim 6-7$ show offset between the UV and IR emission \citep{inami22,bowler22}, suggesting non-uniform dust distributions. However, it remains to be shown whether the shallow slopes observed at low {\AV} in high-redshift galaxies are due to a higher fraction of dust-free sightlines compared to the low-redshift galaxies with similarly low {\AV} values.
Using the inclination and size of our galaxies, we found no evidence that the attenuation curve slope depends on these parameters. However, our data is not sufficient to detect dust distribution variations on parsec scales, which is more relevant for high-$z$ galaxies if their UV continuum emission is dominated by OB associations.
On the other hand, as discussed below, hydrodynamical simulations that include dust evolution provide strong evidence that changes in dust grain properties driven by ISM chemical evolution, can reproduce the observed slope–{\AV} evolution with redshift.

We start with comparing the static and dynamic grain size evolution models described in \S\ref{sec:slope_morph} and shown in Figure~\ref{fig:slope_av_models_inclination}. These are non-cosmological models for an isolated MW-type galaxy.
Compared to the static grain-size model (where grain size distribution and composition are fixed), the dynamic grain-size model (where there is a time evolution of grain properties) shows shallower slopes at a given {\AV}, particularly at early times ($t_{\mathrm{age}}<1$\,Gyr). At early stages ($t_{\mathrm{age}} < 0.25$\,Gyr), small grains ($a < 0.05\,\mu$m) have not yet formed, resulting in lower $A_\mathrm{FUV}$ especially at high inclination angles. Furthermore, large grains ($a > 0.1\,\mu$m), which have high albedo, are also not abundant, suppressing the effect of scattering. As a result, the attenuation curves in the dynamic grain-size model are shallower, consistent with the observations at high redshifts ($z > 1$), suggesting that the attenuation curves predicted by the dynamic grain-size model are more suitable for early galaxies.

While these simulations \citep{matsumoto26} provide valuable insight into the effects of grain size evolution in MW-type galaxies, we also examine another set of simulations spanning a range of galaxy masses and metallicities. We adopt the hydrodynamical simulations in the \texttt{RAMSES} code from \cite{dubois24}, which couple dust growth to the chemical enrichment of the gas. In these simulations, isolated disk galaxies with varying masses, metallicities, and gas fractions are simulated. Dust is categorized into small ($\sim 5$\,nm) and large ($\sim 0.1\,\mu$m) grains, with two main compositions: carbonaceous grains (graphite-like) and silicate grains. Key processes shaping dust evolution include growth via accretion and stellar ejecta, destruction through supernova shocks, thermal sputtering, and astration, and size evolution through shattering (fragmentation of large grains) and coagulation (merging of small grains). The simulations are done for isolated disk galaxies in three categories: a) Milky Way-like systems (G10LG, halo mass of $M_{\mathrm h}=10^{12}\,M_{\odot}$), b) intermediate-mass galaxies ($M_{\mathrm h}=10^{11}\,M_{\odot}$), G9LG with low gas fraction and G9HG with high gas fraction, and c) low-mass galaxies (G8HG, $M_{\mathrm h}=10^{10}\,M_{\odot}$). These galaxies are further modeled with different metallicities (e.g., high-metallicity G10LG-HZ, very low-metallicity G9LG-VLZ). For more detail on the simulations, refer to \cite{dubois24}. Particularly relevant to the results shown in this paper, the grain size distribution of these models are shown in Figure 5 and Appendix E of \cite{dubois24}. The final simulation outputs are then post-processed with \texttt{SKIRT} \citep{baes11,camps15} to compute attenuation curves, using the full dust spatial distribution and properties as predicted by the simulation to perform the radiative transfer calculations. The results of the simulations are shown in Figure~\ref{fig:slope_av_models}. Within each simulation category (shown by different colors in Figure~\ref{fig:slope_av_models}), individual data points represent variations in initial metallicity, dust evolution assumptions, and a range of inclination angles. As before, black contours are data in this study from $z\sim1$ to 9. 

The massive halo simulations (G10, green and red points in Figure~\ref{fig:slope_av_models}) successfully reproduce the average trend observed in the $z\sim 0$ sample: the high metallicity, low gas fraction systems that resemble the local Universe late-type galaxies have relatively low {\AV} values but steep curve slopes. These steep slopes arise from efficient dust growth and processing (shattering) within the ISM, which increases the fraction of small dust grains and consequently leads to high attenuation at short wavelengths. On the other hand, the high-metallicity systems with high gas fractions, typically starburst galaxies, populate the tail characterized by high {\AV} and shallow slopes. This behavior is attributed to increased grain coagulation in these environments, which leads to the growth of larger dust grains \citep{asano13,hirashita15,hirashita19} and results in a flattening of the attenuation curve slope \citep{weingartner01a,hirashita20,ysard24}. This interplay between metallicity and dust grain evolution also explains the previously observed trends seen at $z \sim 2$, where low-metallicity galaxies tend to have steep slopes and higher-metallicity galaxies flatter ones \citep{shivaei20a}.

Interestingly, the lower-mass halo simulations (G8/G9, shown as orange points in Figure~\ref{fig:slope_av_models}) with low metallicity but high gas fractions occupy a similar locus as the high-redshift galaxies in the slope-{\AV} diagram. They have shallower slopes at low {\AV} values compared to the G10 simulations. This behavior is a result of the high gas fractions combined with low metallicities that leads to very inefficient dust growth and metal accretion in the ISM.  
As a result, dust produced by SNe (in their ejecta) dominates the total dust population. As an assumption in the models, the size distribution of dust grains produced by stars is skewed towards large sizes \citep{nozawa03, nozawa07,bianchi07,gall11,gall14,sarangi15,gall18}\footnote{While some theoretical work suggest the opposite of the assumption in this work regarding the size distribution of condensed dust grains in SNe ejecta \citep[see the review of][]{schneider-maiolino24}, the observations \citep[e.g.,][]{gall14,wesson15,priestley20,niculescu22} have been showing strong evidence that the majority of dust grains in SN ejecta are indeed large (about micron-sized).}, leading to a shallower attenuation curve slope. 
It is only when the metallicity reaches a sufficiently high level that small grains begin to form via shattering processes in the ISM, which subsequently coagulate into larger grains within the dense ISM. 

In summary, each model in Figure~\ref{fig:slope_av_models} shows a decreasing trend of slope with {\AV}, driven by variations in dust–star geometry as well as dust properties due to the chemical evolution of the ISM, which influences dust production and processing. In contrast, the {\AV}–slope trend is systematically different between models with different masses and initial metallicities (red and orange symbols), reflecting differences in the origin and evolution of dust that leads to distinct grain size distributions. While scattering and dust–star geometry influence the overall {\AV}–slope relation, it is the grain size distribution that distinguishes low- and high-mass (and metallicity) galaxies in the simulations.

Comparison of the observed decrease in attenuation curve slope with redshift at a given {\AV} to these simulations suggests that dust in high-$z$ galaxies is primarily of stellar origin with limited ISM processing. Although this trend seems in contrast with the steep attenuation curves seen in low-metallicity galaxies at lower redshifts, the key difference lies in the dust origin: at high redshifts, low metallicity does not imply enhanced shattering in the ISM, but rather reflects a lack of efficient dust growth and processing, resulting in dust dominated by large grains formed in stellar ejecta.

\begin{figure*}[!ht]
	\centering
	\includegraphics[width=.43\textwidth]{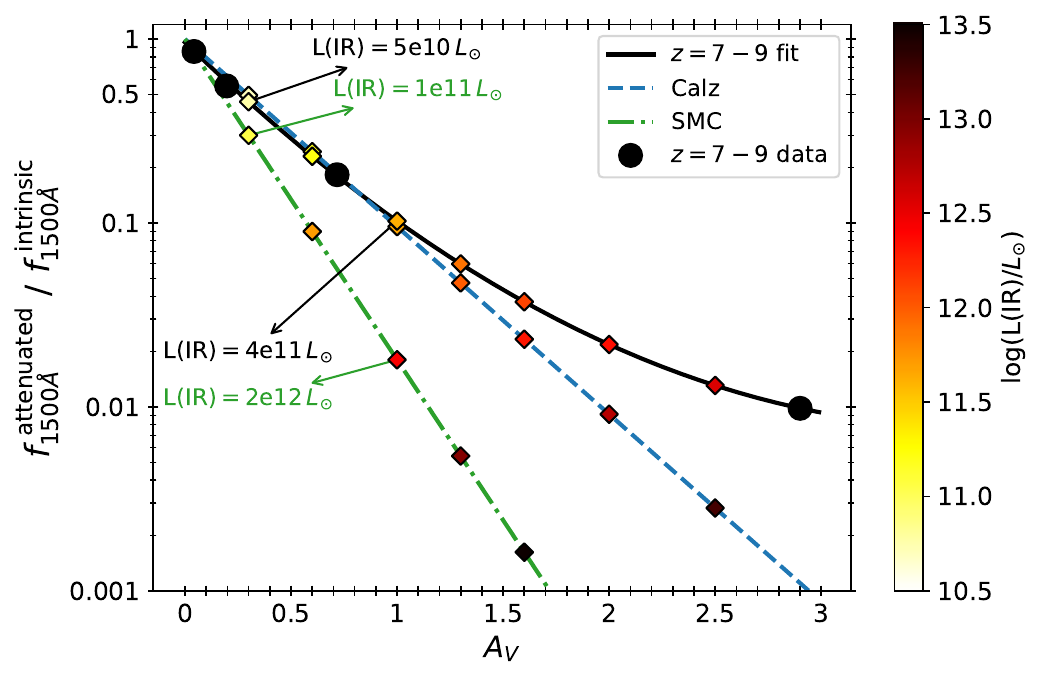}\quad
	\includegraphics[width=.45\textwidth]{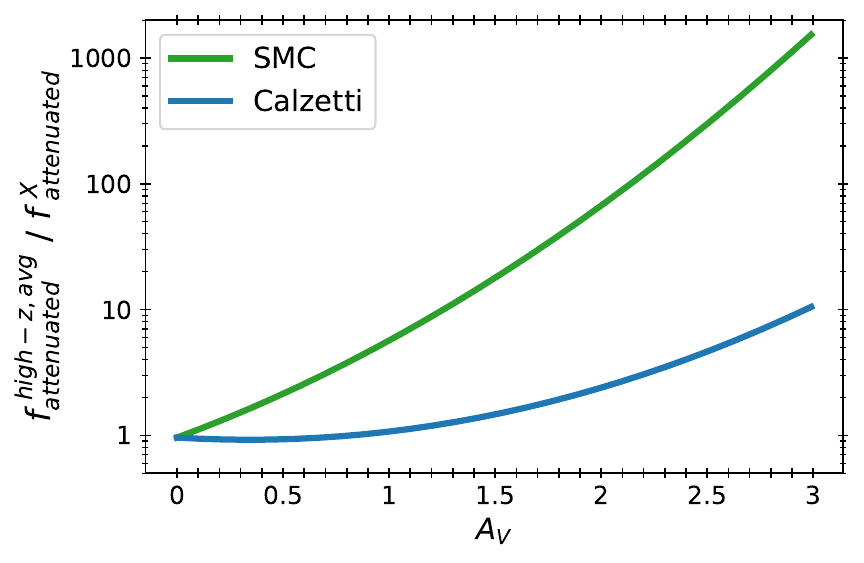}\quad
    
	\caption{Left: ratio of attenuated to intrinsic UV flux at 1500\,$\AA$, or dust transmission fraction, as a function of optical attenuation, {\AV}. The SMC, Calzetti, and high-$z$ average curves are shown. The high-$z$ average curve is derived by fitting a second-order polynomial function to the $A_V-A_{1500}$ relationship of the average curves at $z=7-9$ (including the highest {\AV} bin that was excluded in the previous sections).
    Colored diamonds show total IR luminosities calculated using Equation~\ref{eq:lir} for a galaxy with $M_{1500}=-20$ AB magnitude ($M^*$ at $z\sim 8$, \citealt{donnan23}) at a few $A_V$ values. The values of IR luminosity at $A_V=0.3$ and 1, assuming the SMC curve (green) and the average high-$z$ curve (black) are annotated in the figure. The gray average high-$z$ curve results in significantly less UV obscuration, and consequently lower IR luminosity, compared to the steep SMC curve.
    Right: ratio of UV (1500\,$\AA$) flux attenuated by the average high-$z$ curve to that attenuated by the SMC (green) and Calzetti (blue) curves. The flatter high-$z$ curve results in lower UV attenuation than these two commonly assumed curves.
	}
	\label{fig:highz_att}
\end{figure*}

\subsection{Much less UV attenuation at $z>7$ than previously expected?}

An intriguing result is the shallow attenuation curve observed in high-redshift galaxies. As shown in Figure~\ref{fig:attcurve_av}, the average curve at high redshift is consistently shallower than the SMC curve. The two high-$z$ galaxies that have {\AV} higher than 2 and are excluded in the previous plots show attenuation curves that are even flatter than the Calzetti curve. The average of these two slopes is shown in Figure~\ref{fig:highz_att}. A flat curve at high redshifts means that for the same amount of optical attenuation (or V-band optical depth, or reddening), high-redshift galaxies, on average, have less UV obscuration, and hence, less IR dust emission, than their lower redshift counterparts. This has important implications for studying the dust content of high-redshift galaxies: dust is present, but it attenuates UV light less efficiently than at lower redshifts, which may offer a potential explanation for the so-called ``blue monsters'' without invoking dust-free scenarios \citep{ferrara25a,ferrara25b}. This idea has also been supported in simulations \citep{narayanan25}. As discussed in the previous section, such shallow curves are expected when the dominant dust population originates from stars and has not undergone significant processing in the ISM. Given the young ages and low metallicities of galaxies at $z>7$, this scenario is within expectations\footnote{This should not be confused with the steeper slopes seen in young, low-metallicity galaxies at lower redshifts, whose higher initial metallicity makes them resemble the G10 simulations in Figure~\ref{fig:slope_av_models} more closely. In contrast, high-z galaxies that resemble the G8/9 simulations show relatively shallower slopes compared to their lower-redshift counterparts due to differences in their dust populations. Nevertheless, within the high-z sample, the slope–{\AV} trend is still expected to be influenced by age and dust–star geometry (e.g., as shown with the simulations in Figure~\ref{fig:slope_av_models_inclination}).}.

In Figure~\ref{fig:highz_att}, we show the fraction of attenuated flux at 1500$\AA$ as a function of {\AV}. The black curve represents the second-order polynomial fit to the {\AV}-$A_{1500}$ relation at $z=7-9$ (including the highest {\AV} point shown in the figure). The figure shows the differences in UV obscuration associated with different attenuation curves. As seen in the right panel, the discrepancy in attenuated UV flux between the Calzetti curve and the SMC curve increases rapidly with increasing {\AV}. At $A_V=0.5$, the UV attenuation for the flat curve is $\sim 3\times$ lower than that of the SMC-curve, and this difference exceeds by $>10\times$ for $A_V>1$. These variations in UV obscuration directly translates into differences in the emitted IR luminosity. Regardless of the dust temperature, which affects fluxes at specific far-IR wavelengths, in the optically-thin regime, the total IR luminosity can be estimated using a simple energy balance argument:

\begin{equation} \label{eq:lir}
    L(\rm{IR})  = \gamma \times \nu L_{\nu}(1500)_{attenuated} \times (10^{0.4A_{1500}}-1),
\end{equation}
where $\nu L_{\nu}(\rm{1500})$ is the attenuated UV continuum luminosity at 1500$\AA$, and $\gamma$ is the bolometric correction to convert the absorption at 1500\,$\AA$ to total absorbed UV flux. We adopt $\gamma=2.13$ from the calibrations of \citet{hao11}. As an example, we estimate the IR luminosities for a galaxy with an absolute UV magnitude of $M_{1500} = -20$ ($M^*$ at $z\sim 8$, \citealt{donnan23}) using different attenuation curves, for discrete {\AV} values. These IR luminosities are shown with the color-coding of diamond symbols in the figure. The diamonds show the range of variation in estimated IR luminosities from different curves at a given {\AV}. For example, at $A_V = 0.3$, the estimated IR luminosity is $2.5\times$ higher assuming the steep SMC curve, and it reaches to about an order of magnitude difference at $A_V = 1$ (annotated values in the figure). Such significant discrepancies can have important implications for predicting fluxes in follow-up observations of high-redshift galaxies with FIR/submm facilities such as ALMA \citep[see, for example, ALMA constraints of undetected dust continuum emission at $z > 10$ in][]{zavala24,schouws25,carniani25,witstok26}.

The example above shows a thought experiment with a galaxy of known dust content and a shallow attenuation curve, where the existing dust is less efficient at attenuating UV-optical light compared to a galaxy with the same dust content but a steep attenuation curve (in this thought experiment, the difference in slopes could arise from different dust grain types or different dust–star geometry). This physical implication of varying dust attenuation properties is different from the practical implication of fitting galaxies SEDs with models that assume different attenuation curves. In practice, when fitting SEDs of high-redshift galaxies, adopting an attenuation curve with a different slope also leads to inferring a different {\AV}. As a test, we fit our $z=7-9$ sample with a fixed SMC curve and compared the inferred properties to those obtained with our flexible attenuation model (we limit the sample to 101 galaxies with $\chi^2 < 1$ to avoid uncertainties due to bad fits). For galaxies whose curves were shallower than SMC in the flexible model ($80\%$ of the sample), using a fixed SMC curve overestimates ages and underestimates SFR and {\AV}. On average, SMC-based ages are $6\times$ higher, while SFR and {\AV} are 0.8 and 0.7 times the values inferred with the flexible curve, respectively. In extreme cases, the discrepancies can be as large as {\em two} orders of magnitude. Stellar mass values are not systematically biased but can vary by as much as 1\,dex in either direction (for comparison, in the flexible-curve sample with steep inferred slopes, stellar masses derived with a fixed SMC curve differ by at most 0.3\,dex). 

\section{Summary}\label{sec:summary}

We present a comprehensive analysis of the dust attenuation curve slopes in a sample of $\sim 3600$ galaxies spanning redshifts of $z\sim 1-9$, drawn from three large JWST/NIRCam grism surveys: ALT, FRESCO, and CONGRESS. The grism sample has the advantage of including all spectroscopically confirmed galaxies identified through emission lines, without the pre-selection biases inherent to slit-spectroscopic surveys. This work utilizes homogeneous SED modeling with the \texttt{Prospector} code, leveraging deep HST and JWST/NIRCam medium and wide band photometry, spectroscopic redshifts, and known emission line fluxes from grism spectra to constrain the intrinsic stellar population and, in turn, the UV–optical shape of the stellar dust attenuation curve. 

Our results show that the attenuation curve slope strongly anti-correlates with total dust optical depth ({\AV}), consistent with previous findings but now established across a significantly larger and more diverse high-redshift sample. The curve slope is also anti-correlated with SFR and stellar mass, however, these dependencies become weak once the effect of {\AV} is taken into account (Figure~\ref{fig:slope_av_params}). We find no evidence for significant variation of the curve slope with inclination (axis ratio) or size parameters (Figures~\ref{fig:slope_av_models_inclination}, \ref{fig:slope_size_sigma}). While we do not find evidence for a dependence of curve slope on inclination angle or age inferred from SED fitting, simulations suggest that the slope–{\AV} relation arises from dust scattering properties, dust–star geometry, and stellar population age \citep[e.g.,][]{chevallard13,narayanan18,matsumoto26}.

Importantly, after controlling for {\AV}, we detect a redshift evolution in the attenuation curve shape (Figure~\ref{fig:attcurve_av}): high-redshift galaxies ($z > 7$) show shallower attenuation curves than those at $z<3$, in some cases flatter than the Calzetti curve. A comparison with literature supports this picture, showing that our $z\sim 1-9$ sample has, on average, flatter attenuation curves than the $z\sim 0$ sample of \cite{salim18}. In other words, although the slope–{\AV} relation is preserved at all redshifts, the locus of high-redshift galaxies shifts downward in this diagram (Figure~\ref{fig:slope_av_literature}). We provide an analytic function for the slope of the attenuation curve as a function of {\AV} and redshift that can be implemented in SED fitting codes (Equations~\ref{eq:fit1} and \ref{eq:fit2}).

While at a given redshift, the slope–{\AV} relation originates from the effects of scattering and star-dust geometry, the systematic downward trend of the slope–{\AV} relation from $z \sim 0$ to 9 likely reflects changes in dust production and processing mechanisms with redshift. To test this hypothesis, we compare our results with attenuation curves derived from the hydrodynamical simulations of \cite{dubois24} with post-process radiative transfer calculations. These models predict a downward shift in the slope–{\AV} diagram when moving from massive to lower-mass halo simulations, qualitatively reproducing our observed evolution from $z \sim 0$ to 9 (Figure~\ref{fig:slope_av_models}). In the simulations, this shift is driven by changes in dust grain properties: low-mass, low-metallicity galaxies are dominated by supernova-produced dust rather than ISM-processed dust, resulting in a grain size distribution skewed toward larger grains and, consequently, a shallower UV–optical attenuation curve. Therefore, the comparison between observations and simulations suggests a dominance of large, unprocessed, stellar-origin dust grains in the early universe, prior to significant ISM processing (i.e., grain shattering and coagulation in the ISM).

These findings have direct implications for interpreting the UV and IR emission of high-$z$ galaxies. In particular, shallower attenuation curves indicate that high-redshift dust grains are less efficient at attenuating UV light and, consequently, re-emitting in the IR. In other words, at a given {\AV}, galaxies at $z > 7$ typically show lower UV attenuation and IR emission than those at $z < 3$ (Figure~\ref{fig:highz_att}), due to differences in dust grain properties despite having the same optical attenuation. This may help explain the unexpectedly low dust attenuation and IR detections in some $z > 7$ galaxies, without requiring them to be dust-free. Practically, the results of this work strongly emphasize the importance of moving away from assuming a single attenuation curve across redshifts and galaxy populations; instead, flexible attenuation curve slopes should be adopted in SED fitting, as there is substantial diversity in attenuation curve slopes both within a given redshift and across redshifts. For example, using an SMC attenuation curve for the majority of $z > 7$ galaxies that have shallower curves, can lead to systematically underestimated SFRs, dust optical depths, and overestimated ages -- in extreme cases, by as large as two orders of magnitude. The exact level of uncertainty in each parameter depends on factors such as the galaxy intrinsic properties, wavelength coverage of the data, and other assumptions adopted in the SED modeling.

\begin{acknowledgements}
We thank the referee for their thorough comments on the manuscript.
IS thanks Laura Sommovigo, Desika Narayanan, and Andrea Ferrara for fruitful discussions on the topic.
This work has been funded by the Atracc\'{i}on de Talento Grant No. 2022-T1/TIC-20472 of the Comunidad de Madrid, Spain, as well as the European Research Council (ERC) under the European Union’s Horizon 2020 research and innovation programme (DistantDust, Grant agreement No. 101117541).
KM is a Ph.D. fellow of the Flemish Fund for Scientific Research (FWO-Vlaanderen), supported by Grant 1169822N.
AA acknowledges support by the Swedish research council Vetenskapsr{\aa}det (VR 2021-05559, and VR consolidator grant 2024-02061).
This work has received funding from the Swiss State Secretariat for Education, Research and Innovation (SERI) under contract number MB22.00072, as well as from the Swiss National Science Foundation (SNSF) through project grant 200020\_207349.
JW gratefully acknowledges support from the Cosmic Dawn Center through the DAWN Fellowship. The Cosmic Dawn Center (DAWN) is funded by the Danish National Research Foundation under grant DNRF140.
This work is based on observations made with the NASA/ESA/CSA James Webb Space Telescope. The data were obtained from the Mikulski Archive for Space Telescopes at the Space Telescope Science Institute, which is operated by the Association of Universities for Research in Astronomy, Inc., under NASA contract NAS 5-03127 for JWST. These observations are associated with programs 3516, 3577, 1895, 1180, 1181, 1963, 2561. This research is also in part based on observations made with the NASA/ESA Hubble Space Telescope obtained from the Space Telescope Science Institute, which is operated by the Association of Universities for Research in Astronomy, Inc., under NASA contract NAS 5–26555. 

\end{acknowledgements}

\bibliographystyle{aa}
\bibliography{bibliography}

\appendix
\section{Mock tests for recovering the curve slope} \label{sec:app-mocks}

\begin{figure*}[!bth]
	\centering
	\includegraphics[width=.4\textwidth]{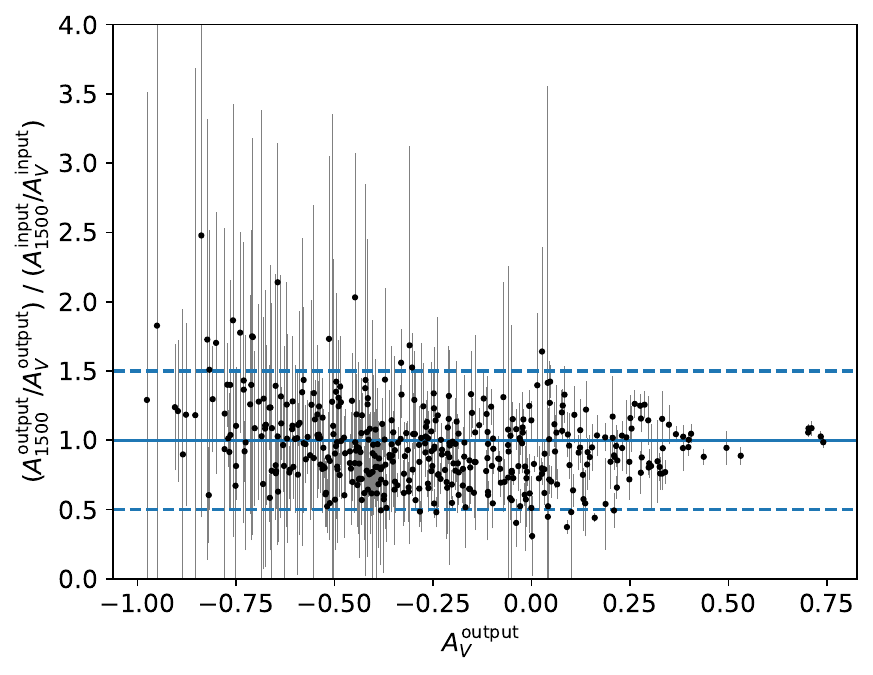} \quad
    \includegraphics[width=.4
    \textwidth]{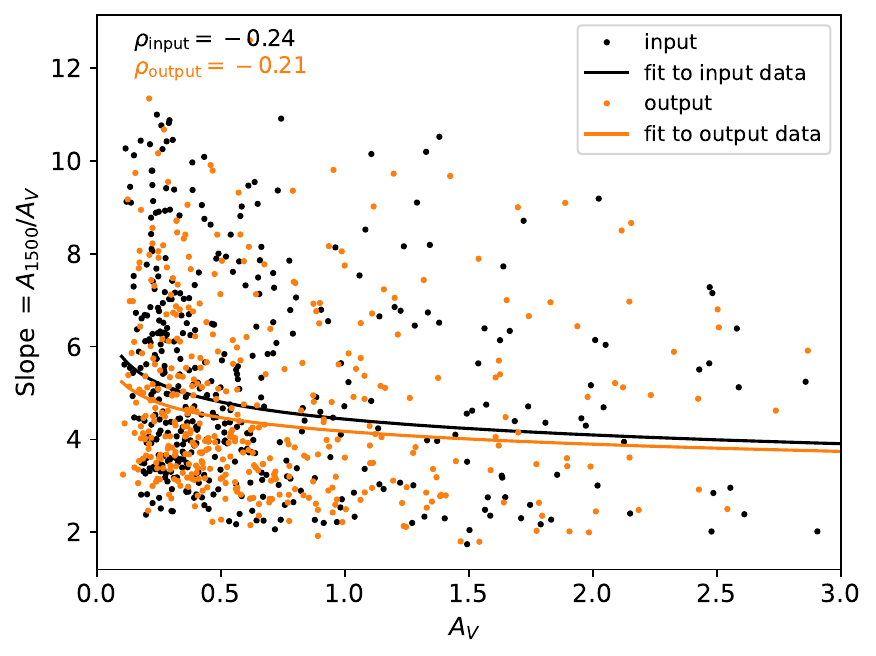}
	\caption{
Mock SED fitting test to evaluate potential degeneracies between the attenuation curve slope and dust optical attenuation.
Left: Comparison between the input and recovered attenuation curve slopes from mock photometry, as a function of {\AV}. While measurement accuracy and precision decrease with lower dust attenuation, the recovered slopes remain consistent with the inputs within their uncertainties, and no significant systematic bias is observed.
Right: Test of whether the observed slope–{\AV} relation could be an artifact of the SED fitting. Black points show the input mock sample, in which slope and {\AV} were randomly assigned and thus uncorrelated. Orange points show the results from fitting the mock data with \texttt{Prospector}, demonstrating that the recovered slope–{\AV} relation preserves the lack of correlation. This confirms that the \texttt{Prospector} fitting procedure does not artificially induce a slope–{\AV} trend.	}
	\label{fig:mocktest}
\end{figure*}

The attenuation curve slope from SED fitting can be degenerate with the dust optical depth ($\tau_V$, or similarly {\AV}). To test for this degeneracy and its potential effect in deriving the slope-{\AV} relation, we perform the following mock SED fitting test.
We randomly select 100 galaxies from the slope–{\AV} parameter space, dividing them into 10 bins of {\AV}. For each {\AV} bin, we define new values for the \texttt{Prospector} \texttt{dust2} parameter ($\tau_V$, the optical depth for old stars) and the \texttt{dust\_index} parameter ($\delta$, the power-law exponent of the attenuation curve in Equation~\ref{eq:kappa}) by randomly sampling within their respective ranges using a uniform prior. This process is repeated five times, generating five new inferences per galaxy. 
In total, we generate 500 new FSPS templates using the newly assigned \texttt{dust2} and \texttt{dust\_index} values, while keeping the other parameters from the original galaxies unchanged. We then create a catalog of mock photometry in the same filters as the original data and fit these mock observations using the same \texttt{Prospector} setup.

First, the left panel of Figure~\ref{fig:mocktest} shows a comparison between the recovered slope values and the original input values. We find that both the accuracy and precision of the slope measurements decrease with decreasing dust optical attenuation ({\AV}), but there is no significant systematic bias present. Overall, the recovered slopes remain consistent with the input values within their uncertainties.

Secondly, we test whether the observed slope–{\AV} relation could be a byproduct of the SED fitting. By design, the mock galaxy sample does not have an intrinsic slope–{\AV} correlation, as the slope and {\AV} values were effectively selected randomly. This is shown with black points in the right panel of Figure~\ref{fig:mocktest}. The fit to the mock galaxies is almost flat, with a negative correlation factor reported in the Figure. If the slope–{\AV} relation were a byproduct of the \texttt{Prospector} SED fitting, we would expect to see a correlation in the outputs of the \texttt{Prospector} fits to the mock data. However, the orange points in the right panel of Figure~\ref{fig:mocktest} show that the recovered slope–{\AV} relation closely matches the input, reflecting the lack of correlation that was built into the mock sample. This confirms that the \texttt{Prospector} fitting does not artificially introduce a slope–{\AV} correlation.

\section{Curves parametrization} \label{sec:curve-fits}
\begin{figure*}[ht]
    \centering
    \includegraphics[width=\textwidth]{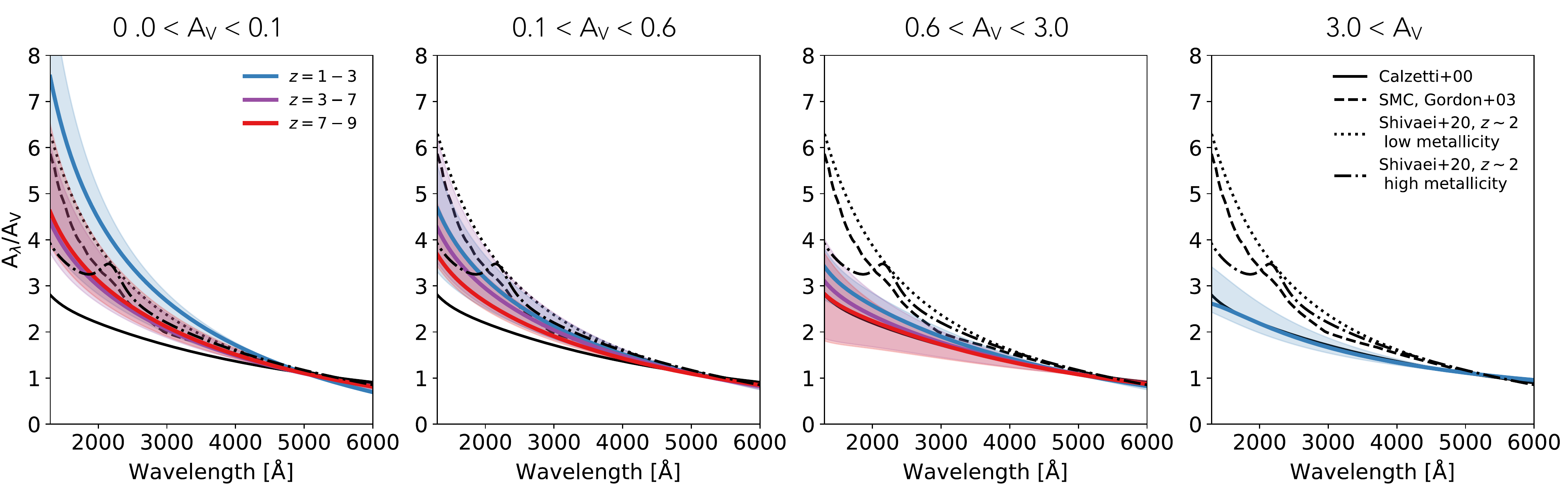}
    \caption{Same as Figure~\ref{fig:attcurve_av}, but showing fits to the data using Equation~\ref{eq:curve-fits}. Fit parameters are shown in Table~\ref{tab:curve-fits}.}
    \label{fig:attcurve_fits}
\end{figure*}

In Table~\ref{tab:curve-fits}, we provide the parameters to functional fits of the attenuation curves presented in Figure~\ref{fig:attcurve_av}. These fits follow the parametrization of UV-optical attenuation curves in \cite{calzetti00}, as:
\begin{equation} \label{eq:curve-fits}
    \frac{A_{\lambda}}{A_{V}} = \alpha_1\left(\frac{1}{\lambda}\right)^3 + \alpha_2\left(\frac{1}{\lambda}\right)^2 + \alpha_3\left(\frac{1}{\lambda}\right) + \alpha_4,
\end{equation} 
where $\lambda$ is in $\mu$m. As the curves at $A_V>2$ are very flat, we use a second-order polynomial with $\alpha_1=0$. The fits are shown in Figure~\ref{fig:attcurve_fits}.

These equations are valid for the UV-optical part of the stellar attenuation curve, and can be combined with the longer-wavelength equation of the Calzetti attenuation curve to provide a full UV-near-IR stellar attenuation curve. For that, an assumption for $R_V$ has to be made (Equation~\ref{eq:absolute}; for example, the Calzetti curves are parameterized in $\kappa_{\lambda}$ with $R_V=4.05$).

\begin{table}[ht]
\caption{Fit parameters for the attenuation curves in bins of {\AV} and redshift shown in Figure~\ref{fig:attcurve_av}, corresponding to those in Equation~\ref{eq:curve-fits}. }
\centering
\begin{tabular}{ccccccc}
\hline\hline  
$A_V$ & $z$ & & $\alpha_1$ & $\alpha_2$ & $\alpha_3$ & $\alpha_4$ \\
\hline
{$0-0.1$} & {$7-9$} 
  & 50th & $0.007$ & $-0.124$ & $1.260$ & $-0.977$ \\ 
  & & 16th & $0.005$ & $-0.093$ & $0.993$ & $-0.559$ \\ 
  & & 84th & $0.009$ & $-0.127$ & $1.472$ & $-1.380$ \\ 
{} & {$3-7$} 
  & 50th & $0.007$ & $-0.120$ & $1.209$ & $-0.895$ \\ 
  & & 16th & $0.006$ & $-0.101$ & $1.001$ & $-0.549$ \\ 
  & & 84th & $0.009$ & $-0.141$ & $1.517$ & $-1.437$ \\ 
{} & {$1-3$} 
& 50th & $0.008$ & $-0.109$ & $1.581$ & $-1.674$ \\ 
& & 16th & $0.005$ & $-0.106$ & $1.108$ & $-0.735$ \\ 
& & 84th & $0.009$ & $-0.067$ & $1.462$ & $-1.603$ \\ 
 \hline\\
{$0.1-0.6$} & {$7-9$} 
  & 50th & $0.007$ & $-0.121$ & $1.113$ & $-0.701$ \\ 
  & & 16th & $0.005$ & $-0.098$ & $0.944$ & $-0.446$ \\ 
  & & 84th & $0.009$ & $-0.151$ & $1.421$ & $-1.211$ \\ 
{} & {$3-7$} 
  & 50th & $0.007$ & $-0.127$ & $1.227$ & $-0.905$ \\ 
  & & 16th & $0.006$ & $-0.112$ & $1.031$ & $-0.575$ \\ 
  & & 84th & $0.009$ & $-0.140$ & $1.512$ & $-1.427$ \\ 
{} & {$1-3$} 
  & 50th & $0.006$ & $-0.107$ & $1.203$ & $-0.911$ \\ 
  & & 16th & $0.006$ & $-0.119$ & $1.071$ & $-0.620$ \\ 
  & & 84th & $0.010$ & $-0.160$ & $1.595$ & $-1.510$ \\ 
 \hline\\
{$0.6-3$} & {$7-9$} 
  & 50th & $0.007$ & $-0.134$ & $1.048$ & $-0.539$ \\ 
  & & 16th & $0.005$ & $-0.102$ & $0.692$ & $0.048$ \\ 
  & & 84th & $0.004$ & $-0.088$ & $0.959$ & $-0.512$ \\ 
{} & {$3-7$} 
  & 50th & $0.007$ & $-0.124$ & $1.032$ & $-0.535$ \\ 
  & & 16th & $0.006$ & $-0.109$ & $0.740$ & $-0.021$ \\ 
  & & 84th & $0.007$ & $-0.134$ & $1.245$ & $-0.911$ \\ 
{} & {$1-3$} 
  & 50th & $0.009$ & $-0.157$ & $1.266$ & $-0.888$ \\ 
  & & 16th & $0.004$ & $-0.099$ & $0.907$ & $-0.371$ \\ 
  & & 84th & $0.013$ & $-0.227$ & $1.687$ & $-1.491$ \\ 
 \\ \hline\\
{$>3$} & {$1-3$} 
  & 50th & -- & $-0.034$ & $0.598$ & $0.055$ \\ 
  & & 16th & -- &  $-0.021$ & $0.440$ & $0.315$ \\ 
  & & 84th & -- &  $-0.038$ & $0.771$ & $-0.278$ \\ 
\hline
\end{tabular}
\label{tab:curve-fits}
\tablefoot{In each {\AV}-redshift bin, the parameters for the 50th, 16th, and 84th percentile fits are shown (Figure~\ref{fig:attcurve_fits}). Refer to Section~\ref{sec:slope_redshift} for details.}
\end{table}

\end{document}